\documentclass[final]{cvpr}

\usepackage{times}
\usepackage{epsfig}
\usepackage{amsmath}
\usepackage{amssymb}
\usepackage{mathrsfs}
\usepackage{graphicx}

\usepackage{float}
\usepackage{booktabs}
\usepackage{multirow}
\usepackage{adjustbox}
\usepackage{subcaption}
\usepackage{transparent}
\usepackage[accsupp]{axessibility}  
\usepackage{nopageno} 

\usepackage[english, main=english]{babel}
\usepackage[pagebackref=true,breaklinks=true,colorlinks,bookmarks=false]{hyperref}

\def\cvprPaperID{9} 
\def\confYear{CVPR 2022}

\begin{document}

\title{Continual Hippocampus Segmentation with Transformers}

\author{Amin Ranem, Camila Gonz\'alez, Anirban Mukhopadhyay\\
GRIS, Technical University of Darmstadt\\
Karolinenpl. 5, 64289 Darmstadt, Germany\\
{\tt\small \{amin.ranem, camila.gonzalez, anirban.mukhopadhyay\}@gris.informatik.tu-darmstadt.de}
}
\maketitle

\begin{abstract}
In clinical settings, where acquisition conditions and patient populations change over time, continual learning is key for ensuring the safe use of deep neural networks. Yet most existing work focuses on convolutional architectures and image classification. Instead, radiologists prefer to work with segmentation models that outline specific regions-of-interest, for which Transformer-based architectures are gaining traction. The self-attention mechanism of Transformers could potentially mitigate catastrophic forgetting, opening the way for more robust medical image segmentation. In this work, we explore how recently-proposed Transformer mechanisms for semantic segmentation behave in sequential learning scenarios, and analyse how best to adapt continual learning strategies for this setting. Our evaluation on hippocampus segmentation shows that Transformer mechanisms mitigate catastrophic forgetting for medical image segmentation compared to purely convolutional architectures, and demonstrates that regularising ViT modules should be done with caution.
\end{abstract}

\section{Introduction}
While Continual Learning (CL) research focuses on image classification benchmarks, the importance of preventing forgetting for segmentation models in clinical settings is being noticed by both medical practitioners and regulatory entities \cite{vokinger2021continual}. Medical data such as Computer Tomography (CT) scans and Magnetic Resonance Images (MRIs) are particularly susceptible to domain shifts caused by changing acquisition parameters \cite{castro2020causality}. Yet privacy regulations set strict constraints on data availability which often make training a model sequentially the only option. As expected, this results in \emph{Catastrophic Forgetting} for early tasks \cite{gonzalez2020wrong,memmel2021adversarial,pianykh2020continuous}.



Transformer-based architectures \cite{dosovitskiy2020image, vaswani2017attention} are gaining traction in medical imaging for semantic segmentation \cite{chen2021transunet, hatamizadeh2021unetr, karimi2021convolution, valanarasu2021medical, wang2021transbts}. Their ability to combine sequential input data with global knowledge by using self-attention \cite{strudel2021segmenter, zheng2021rethinking} could potentially mitigate forgetting. Yet the application of Transformers in medical imaging is not focused on continual segmentation and the majority of CL research focuses on (1) convolutional-based architectures and (2) image classification \cite{delange2021continual, hadsell2020embracing}. In this article, we analyse how susceptible Transformer-based segmentation architectures are to continual shifts in the data distribution, and explore whether self-attention mechanisms can be leveraged for continual segmentation.

The majority of recent work on semantic segmentation \cite{baweja2018towards, mcclure2018distributed, ozgun2020importance,van2019towards, zhang2021comprehensive} consists of regularisation-based techniques in combination with convolutional architectures to reduce overfitting \cite{TEUWEN2020481}.

We propose a backbone model that combines state-of-the-art medical segmentation with Transformers. Additionally, we examine whether popular regularisation-based methods improve the performance when used in combination with Transformers. We augment the state-of-the-art nnU-Net segmentation framework \cite{isensee2018nnu} with Transformer techniques that leverage the self-attention mechanism for medical image segmentation. More specifically, we propose an architecture where the Vision Transformer (ViT) is placed in between the encoding and decoding blocks of the U-Net, and the skip connections are used to build the input of the Transformer architecture (ViT U-Net).

\begin{table}[htb]
\begin{center}
\begin{adjustbox}{max width=\linewidth}
{\begin{tabular}{lcc}
\toprule
& \multirow{2}{*}{\shortstack{Segmentation \\Performance}} & \multirow{2}{*}{\shortstack{CL \\Performance}} \\ 
&& \\ \midrule \midrule
nnU-Net   & \textbf{+ +} & \textbf{-- --} \\
ViT       & \textbf{-- --} & \textbf{+ --} \\
ViT U-Net & \textbf{+ +} & \textbf{+ +} \\ \bottomrule
\end{tabular}}
\end{adjustbox}
\caption{Comparison in terms of segmentation and CL performance. '\textbf{+ +}' means that the method is perfectly suited for this category, '\textbf{+ --}' that it is partially suitable '\textbf{-- --}' that it is not.}
\label{tab:qual_tab}
\end{center}
\end{table}

Table \ref{tab:qual_tab} compares the state-of-the-art nnU-Net with the ViT and our proposed ViT U-Net in terms of segmentation and CL performance. It shows that the proposed ViT U-Net is well suited for both categories.

We analyse the impact of freezing and regularising certain components within the context of Catastrophic Forgetting. The results show that freezing and regularising architecture components can have a positive influence on the amount of maintained knowledge if applied with caution.

We perform our evaluation on the problem of hippocampus segmentation, which is of utmost importance for diagnosing and selecting a promising treatment for neuropsychiatric disorders \cite{carmo2021hippocampus} yet very susceptible to distribution shifts \cite{sanner2021reliable}. Our results show that self-attention can be leveraged to maintain knowledge in a CL setup, whereas regularising the ViT attention module has a negative effect and should be applied with caution. To summarise, our contributions are two-fold:
\begin{enumerate}
    \item we analyse how the problem of Catastrophic Forgetting manifests in Transformer vs. convolutional-based U-Net architectures, and
    \item we propose two variations of our ViT U-Net backbone model which leverages self-attention to mitigate Catastrophic Forgetting for medical image segmentation.
\end{enumerate}

\section{Methods}
\begin{figure}[htb!]
\begin{center}
  \includegraphics[trim={6cm 0cm 6cm 0cm},clip,width=\linewidth]{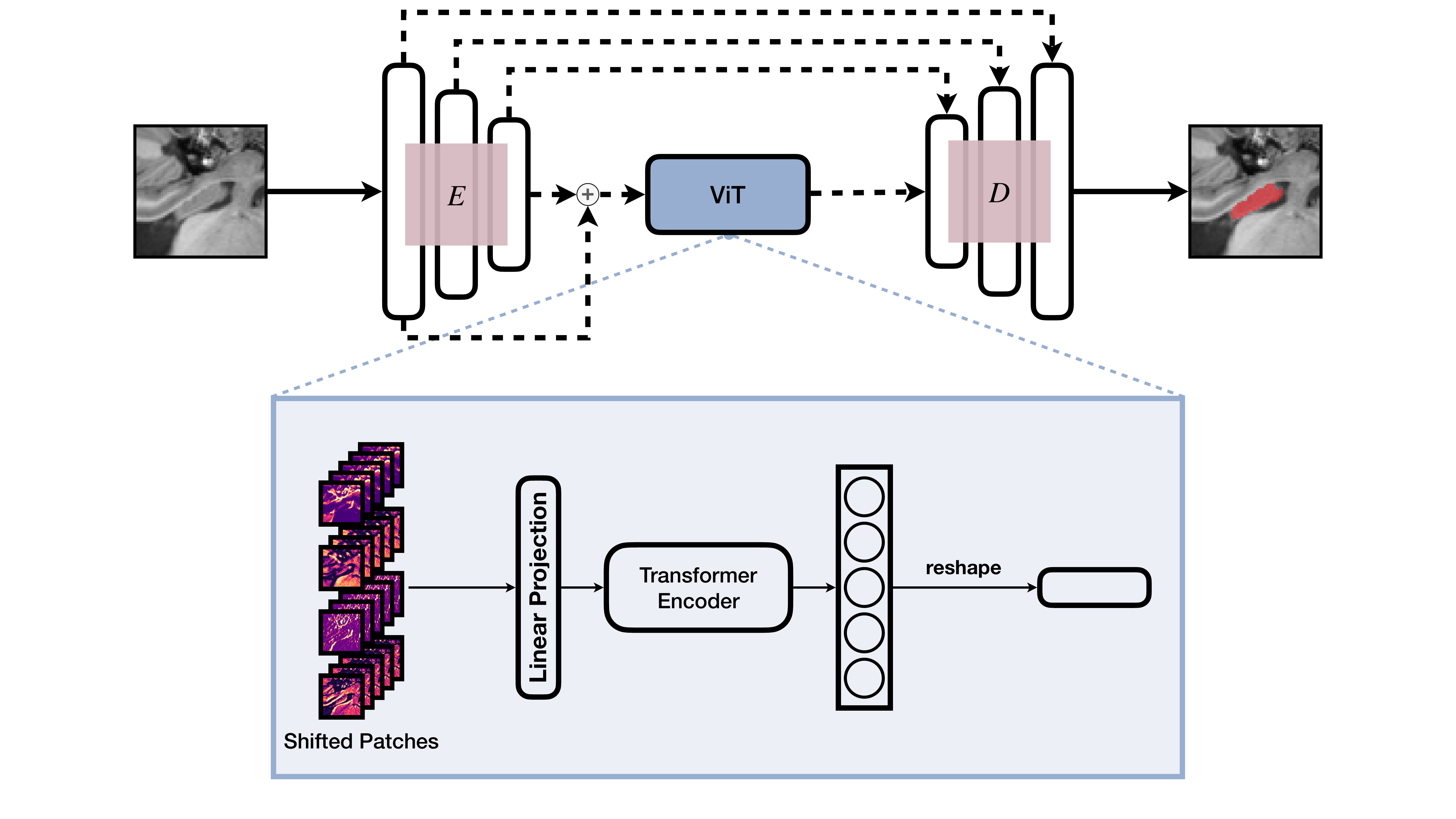}
\end{center}
  \caption{Composition of the nnU-Net and ViT, our proposed ViT U-Net V2. $E$ indicates the encoding and $D$ the decoding blocks of the nnU-Net.}
\label{fig:vit_arch}
\end{figure}

We first present our proposed ViT U-Net which is the composition of the well known nnU-Net \cite{isensee2018nnu} and ViT \cite{dosovitskiy2020image} into one self-adapting architecture, visualised in Figure \ref{fig:vit_arch}. Secondly, we outline how we freeze certain architecture components and apply regularisation to assess the role of the ViT for CL. We define our problem setting as a classic \emph{sequential} training setup, where a network is trained sequentially on a sequence of tasks while performing transfer learning. Finally, we briefly outline CL methods that we explore in combination with our proposed architecture.

\subsection{ViT U-Net}
With the recent publication of the ViT architecture by Dosovitskiy \etal~\cite{dosovitskiy2020image}, new interesting possibilities are opened to leverage the self-attention mechanism known from Transformers for Computer Vision tasks. Within this context, the ViT can potentially be used for medical image segmentation to \textbf{increase the amount of maintained knowledge in a CL setup} as the attention mechanism is designed to access and build upon previously seen states \cite{strudel2021segmenter, zheng2021rethinking}. For this purpose, we propose two different versions of the ViT U-Net, a composition of the self-adapting nnU-Net framework and the ViT architecture, where the ViT is placed in between the encoding and decoding blocks of the U-Net (see Figure \ref{fig:vit_arch}). 

The \emph{U-Net} architecture proposed by Ronneberger \etal~\cite{ronneberger2015u} consists of encoding and decoding blocks -- \textit{indicated in the shape of a U} -- which gives the architecture its prominent name. Between these sections, skip connections ensure that spatial information is maintained.

The nnU-Net is a dynamic framework that performs relevant pre- and post processing steps and adapts the architecture and training configuration of U-Net models based on characteristics of the training data. The framework is state-of-the-art for several medical segmentation challenges \cite[Figure 1]{isensee2019nnu}. 

As mentioned by Chen \etal~\cite{chen2021transunet}, the use of intermediate convolutional feature maps as input of the ViT yields better results as opposed to raw input images. For this purpose, the nnU-Net skip connections are used as the input of the ViT. In addition, we solely focus on skip connections in order to keep the amount of resource allocation as low as possible. Particularly in medical imaging this plays a key role in the applicability of the models due to the large dimensionality of CT and MRI scans.

We present two versions of our proposed architecture where we augment an nnUNet with ViT modules:
\begin{itemize}
    \item The high-level version (V1) considers only high-level features, \ie the very first skip connection is used to build the ViT dimensions and thus used as input for the ViT.
    \item The all-level version (V2) on the other hand considers both high- and low-level features by combining the first and last skip connection using transposed convolutional layers. Based on the findings of Hua \etal~\cite{hua2018lahnet}, those extracted features are combined through a channel-wise addition as shown in Figure \ref{fig:vit_arch} symbolised with the $\oplus$ symbol.
\end{itemize}

Recent work by Lee \etal~\cite{lee2021vision} introduced Shifted Patch Tokenization (SPT) and Locality Self-Attention (LSA) to enable an efficient training on small size datasets when using ViT. We also explore this strategy in our ablation study. 

\subsection{Role of ViT for Continual Learning}
\label{ssec:frozen_setup}
To analyse the role of the ViT in the ViT U-Net, we conduct two different but similarly composed analyses.

\begin{figure}[htb!]
  \begin{center}
  \begin{subfigure}{.9\linewidth}
  \includegraphics[trim={17.5cm 10.5cm 17.5cm 10cm},clip,width=\linewidth]{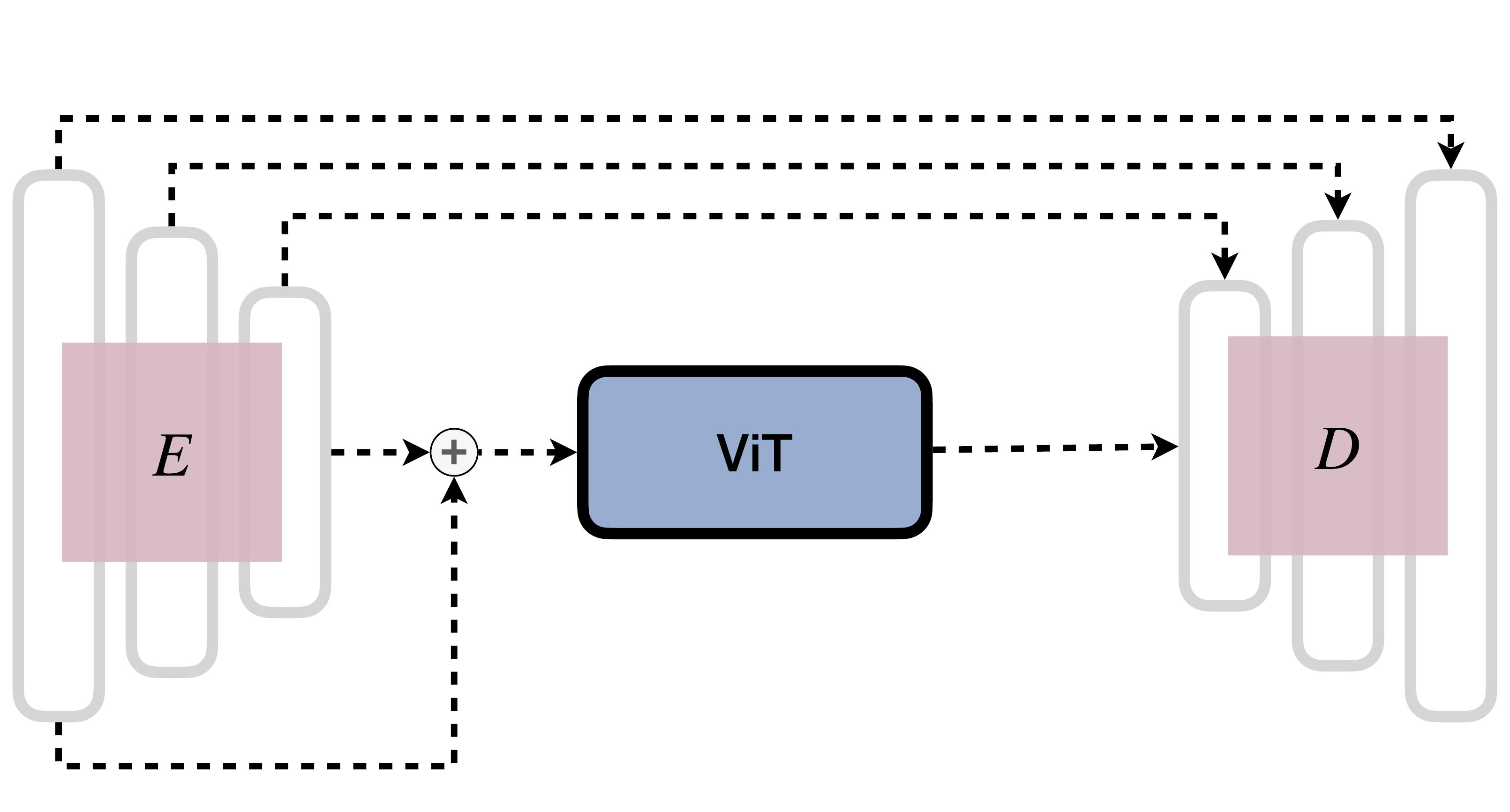}
  \caption{ViT U-Net architecture with frozen U-Net.}
  \end{subfigure}
  \hspace{1 cm}
  \begin{subfigure}{.9\linewidth}
  \includegraphics[trim={17.5cm 10.5cm 17.5cm 10cm},clip,width=\linewidth]{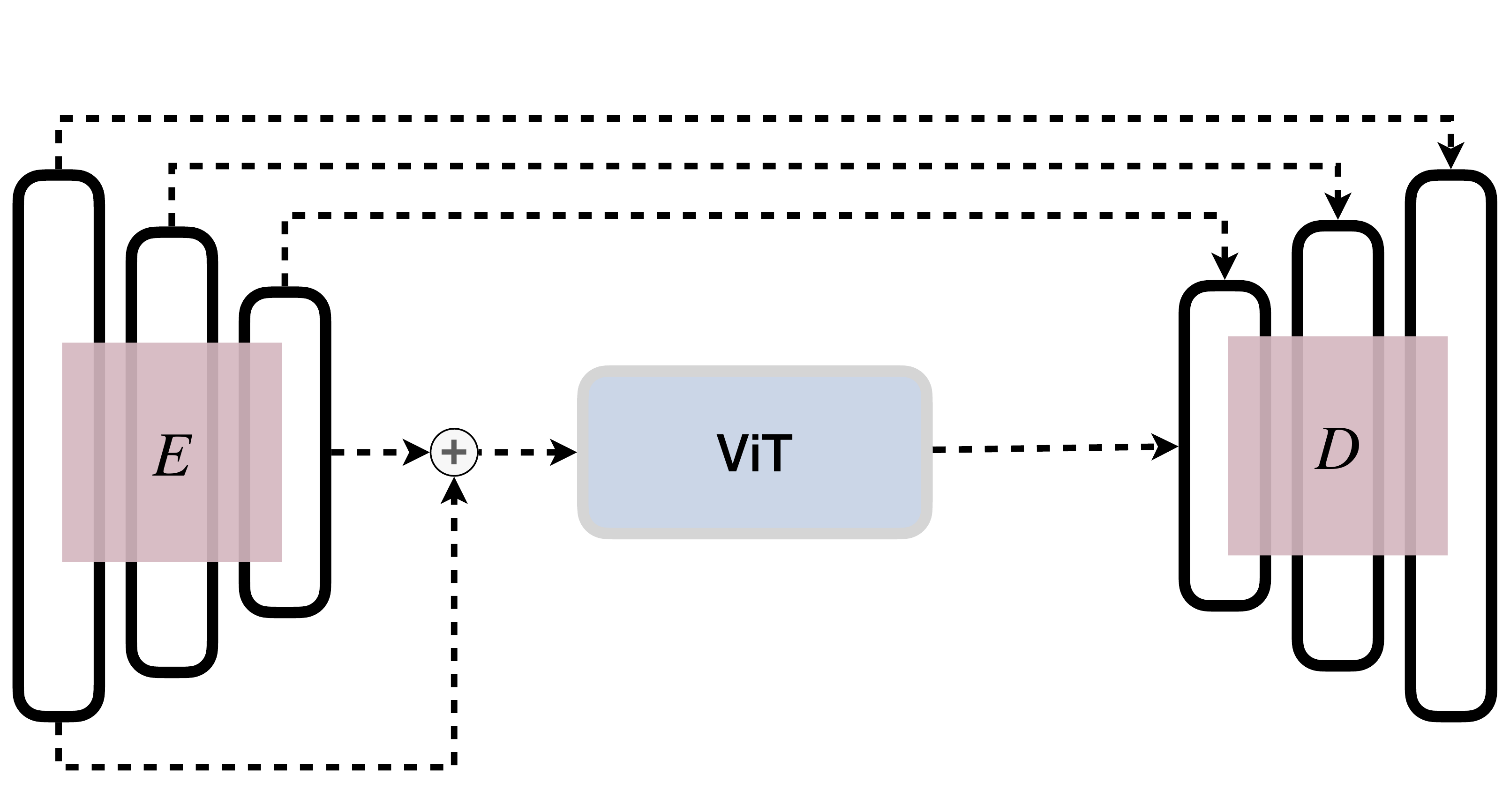}
  \caption{ViT U-Net architecture with frozen ViT.}
  \end{subfigure}
  \caption{ViT U-Net V2 architecture with frozen parts represented in grey. $E$ indicates the encoding and $D$ the decoding blocks of the nnU-Net.}
  \label{fig:frozen_arch}
  \end{center}
\end{figure}

In our first analysis we consider freezing different components of the ViT U-Net. This helps to understand the purpose and benefit of using Transformer-based architectures for medical image segmentation. Models are trained in a sequential setup, where we freeze the specific components after training on the first dataset, as illustrated in Figure \ref{fig:frozen_arch}. With this method, we gain insights on which components are most relevant for learning new patterns and which ones are used to maintain previous knowledge.

Following the same structure, we apply regularisation on specific architecture components during the training process. Among other strategies, we use the well known Elastic Weight Consolidation \cite{kirkpatrick2017overcoming}. With the help of this analysis we want to learn how the regularisation loss affects the process of maintaining knowledge as well as the impact on Transformer-based architectures. To indicate that regularisation is applied on a specific architectural part \texttt{<arch>}, we use the $\mathcal{E} : \text{\texttt{<arch>}}$ notation. When it is applied on the whole network, a simple $*$ is used instead.

\subsection{Continual Learning methods}
We explore the following continual learning methods. As a replay-based strategy, we use simple \emph{rehearsal} which interleaves $25 \%$ of the data randomly selected from every previous task the network has been trained on. 

In terms of regularisation-based methods, we explore the popular \emph{Elastic Weight Consolidation (EWC)} \cite{kirkpatrick2017overcoming}, which makes use of the Fisher Information Matrix $\mathcal{F}$ to ensure that important parameter values are not heavily modified. We also consider \emph{Riemannian Walk (RWalk)} \cite{chaudhry2018riemannian}, which combines EWC with Path Integral to measure parameter importance, but calculates $\mathcal{F}$ in an online fashion throughout the training process instead of after training with each task.

\emph{Modeling the Background (MiB)} \cite{cermelli2020modeling} is a method specifically developed for incremental learning combining knowledge distillation and a modified Cross Entropy Loss that should tackle the \emph{Background Shift} phenomenon in semantic segmentation. The \emph{Pseudo-labeling and Local Pod (PLOP)} \cite{douillard2021plop} falls into the same category as MiB, however it combines a multi-scale spatial distillation loss with pseudo-labeling to increase the amount of maintained knowledge. Based on this method, we consider the \emph{Pooled Outputs Distillation (POD)} method as well. Unlike the POD-Net presented by Douillard \etal~\cite{douillard2020podnet}, we follow the same principle as PLOP but only apply the spatial POD embedding as distillation loss while pseudo-labeling is omitted. We only focus on the spatial POD embedding to understand the importance of pseudo-labeling for medical imaging.

\begin{figure}[htb]
\begin{subfigure}[t]{.49\linewidth}
\centering
\includegraphics[trim={10cm 2cm 10cm 2cm},clip,width=\linewidth]{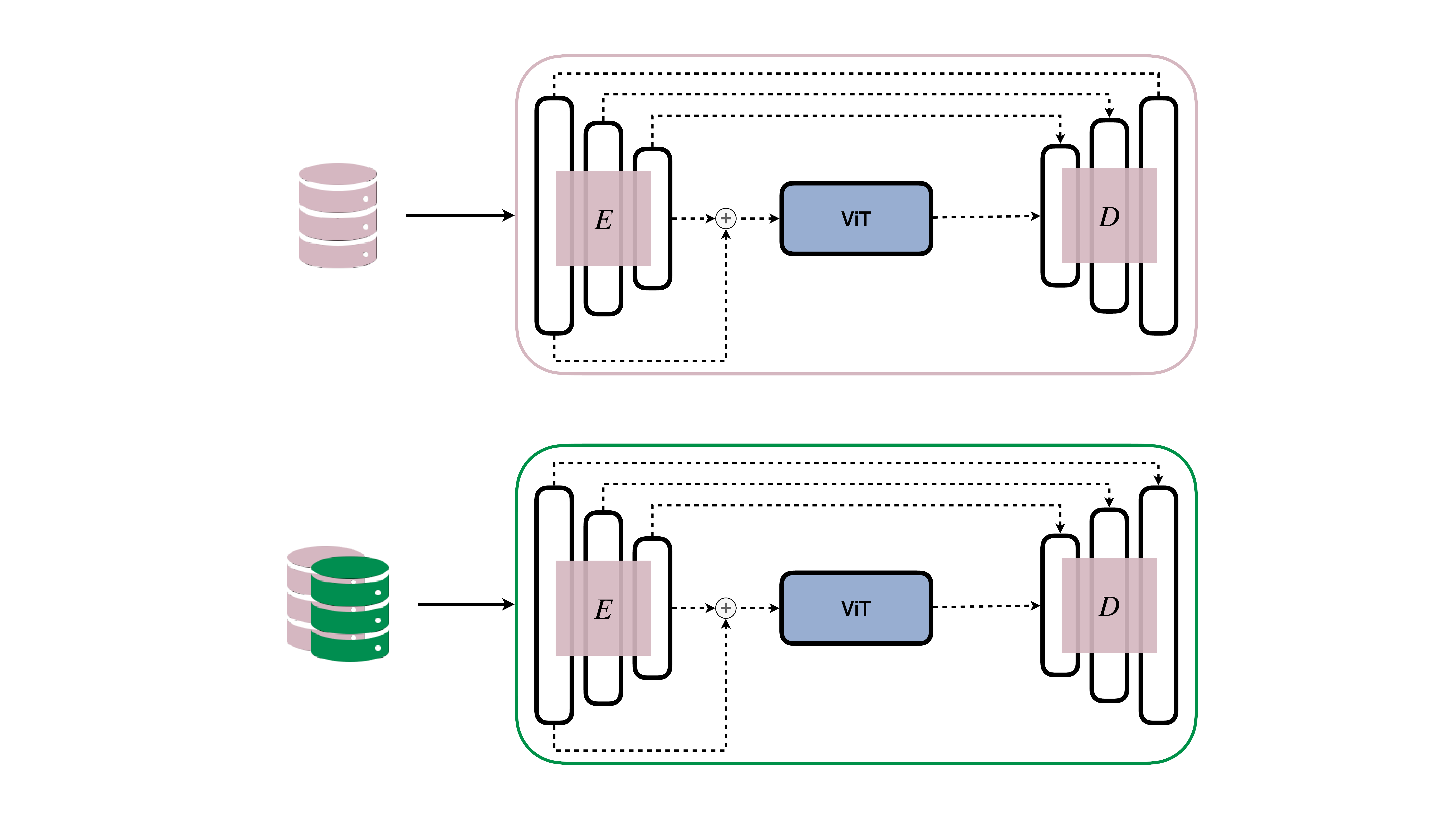}
\caption{Replay-based approach.}
\label{fig:rep}
\end{subfigure}%
\begin{subfigure}[t]{.49\linewidth}
\centering
\includegraphics[trim={10cm 2cm 10cm 2cm},clip,width=\linewidth]{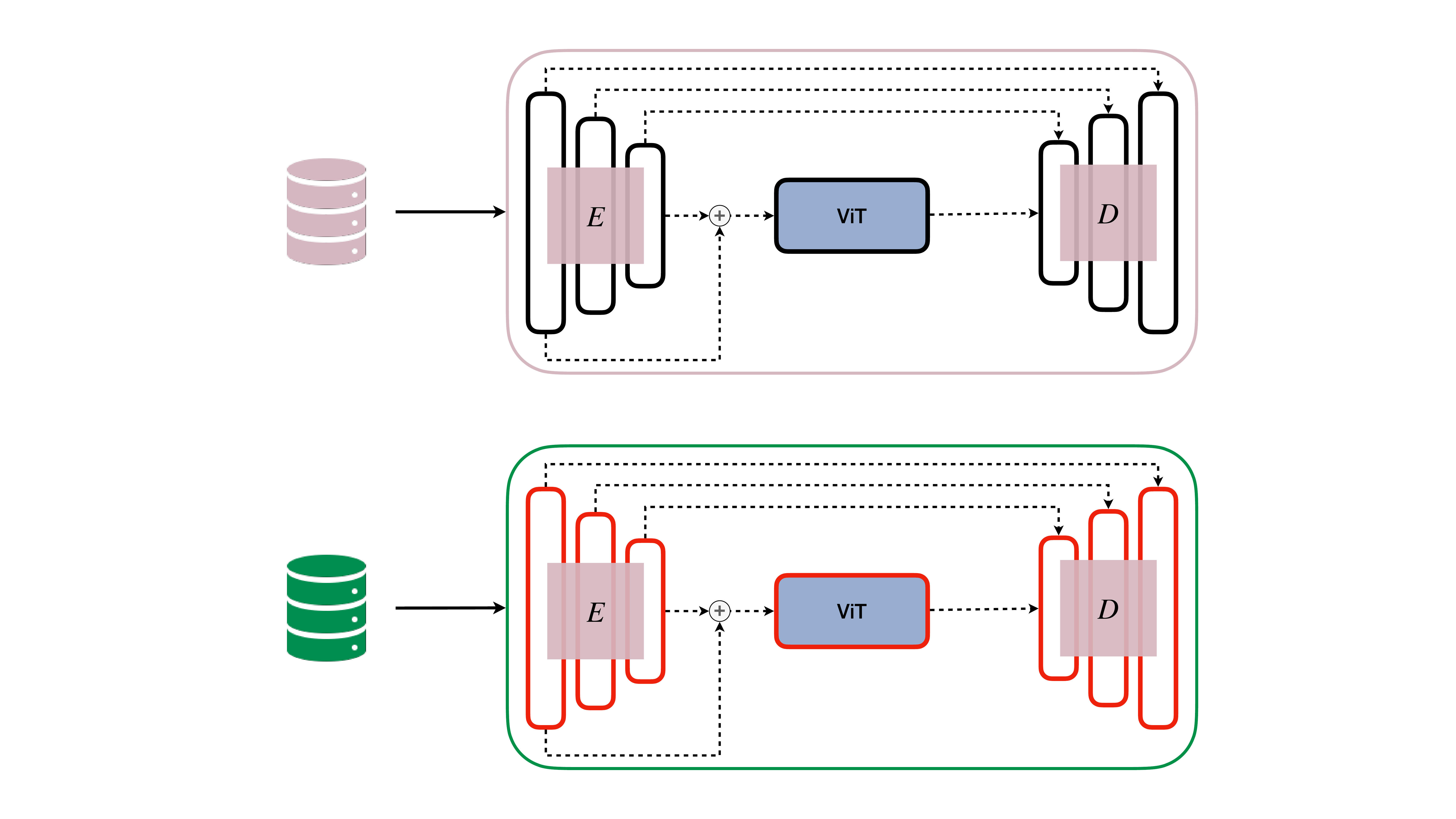}
\caption{Regularisation (red) as in RWalk and EWC \cite{chaudhry2018riemannian, kirkpatrick2017overcoming}.}
\label{fig:reg}
\end{subfigure}\\[1ex]
\begin{subfigure}{\linewidth}
\centering
\includegraphics[trim={6cm 13cm 6cm 3cm},clip,width=\linewidth]{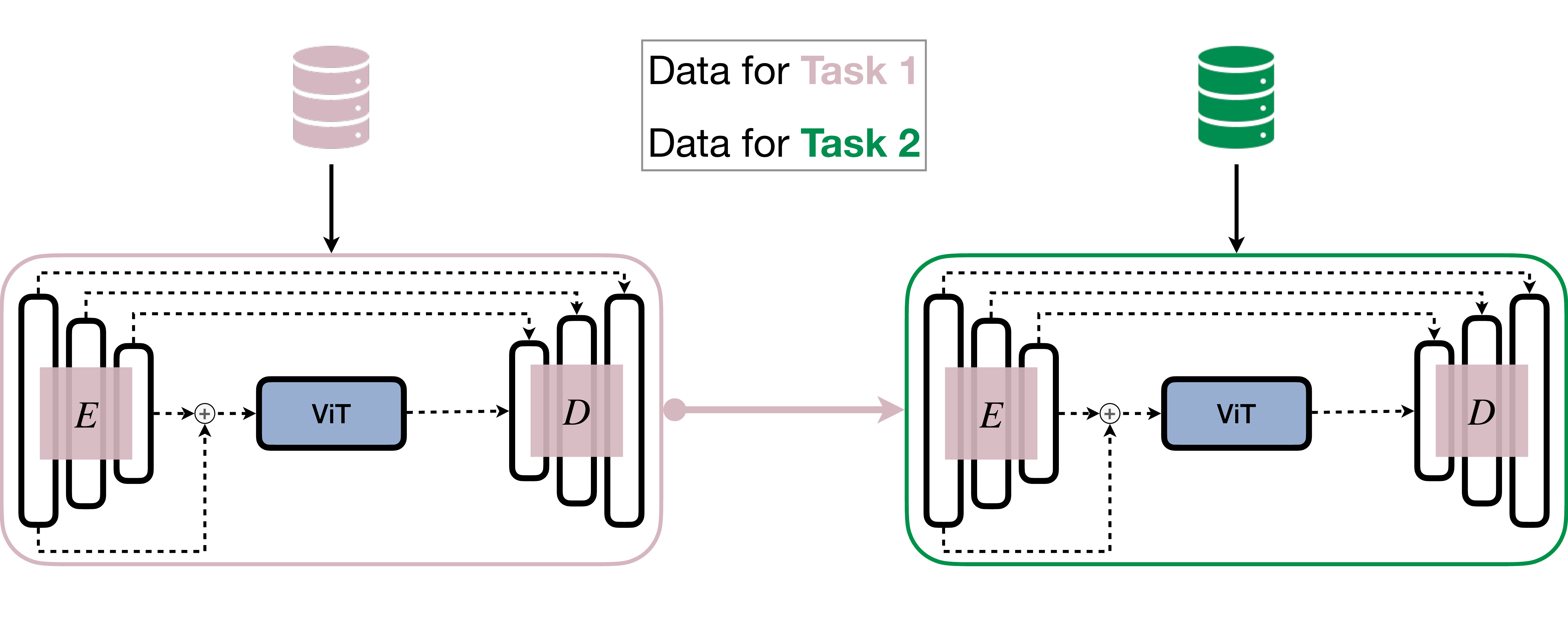}
\caption{Knowledge Distillation as in MiB, PLOP and POD \cite{cermelli2020modeling, douillard2021plop, douillard2020podnet}.}
\label{fig:KD}
\end{subfigure}
\caption{Existing CL methods.}
\label{fig:cl_m}
\end{figure}

Figure \ref{fig:cl_m} demonstrates the different existing CL methods used in this work. In general, replay-based methods interleave data from previous tasks as shown in Figure \ref{fig:rep}. As mentioned previously, regularisation-based methods ensure that certain parameters are not modified or only to a certain degree as done by Kirkpatrick \etal~\cite{kirkpatrick2017overcoming} or Chaudhry \etal~\cite{chaudhry2018riemannian}. Within this context, Figure \ref{fig:reg} shows the general idea behind regularising different layers/parameters during training. Knowledge distillation is commonly used as well, \eg in the MiB \cite{cermelli2020modeling}, PLOP \cite{douillard2021plop} or POD \cite{douillard2020podnet} method. Figure \ref{fig:KD} visually demonstrated the concept behind it.

\section{Experimental Setup}
We start by briefly presenting our dataset corpus. Afterwards, we introduce our hardware settings, training and parameter selection process.

\subsection{Hippocampus data corpus}
Our data corpus contains a total of three hippocampus segmentation datasets, each consisting of T1-weighted MRI scans. The \textit{Harmonized Hippocampal Protocol} dataset \cite{boccardi2015training}, which we refer to as \textit{HarP}, contains healthy subjects and patients with Alzheimer's disease. The second dataset, \textit{Dryad} \cite{kulaga2015multi}, contains 50 cases of healthy patients. Lastly, the hippocampus data from the Medical Decathlon Challenge \cite{medicaldecathlon}, henceforth referred to as \textit{DecathHip}, comprises cases of healthy and schizophrenia patients. As the datasets are annotated with either one or two labels which are not consistent throughout the corpus, we align label characteristics by joining them into one single \emph{hippocampus} class indicating both the posterior and anterior regions of the hippocampus. Table \ref{tab:data} gives an overview of the different dataset characteristics, among others the mean resolution and voxel spacing.

\begin{table}[htp]
\centering
\begin{adjustbox}{max width=\linewidth}
{\begin{tabular}{lcccc}
\toprule
\multirow{2}{*}{Dataset} & \multirow{2}{*}{\shortstack{\# Cases \\ {(train, val)}}} & \multirow{2}{*}{Resolution} & \multirow{2}{*}{Spacing} & \multirow{2}{*}{Source} \\
& & & \\
\midrule \midrule
HarP & 270 -- (216, 54) & [48 64 64] &  [1.00 1.00 1.00] & \cite{boccardi2015training}\\
Dryad & 50 -- (40, 10) & [48 64 64] &  [1.00 1.00 1.00] & \cite{kulaga2015multi}\\
DecathHip & 260 -- (208, 52) & [36 50 35] &  [1.00 1.00 1.00] & \cite{medicaldecathlon}\\
\bottomrule\\
\end{tabular}}
\end{adjustbox}
\caption{Characteristics of our hippocampus data corpus; including number of cases, resolution and voxel spacing.}
\label{tab:data}
\end{table}

\begin{figure}[htp]
\begin{center}
   \includegraphics[clip, trim=13cm 7cm 13cm 7cm, width=\linewidth]{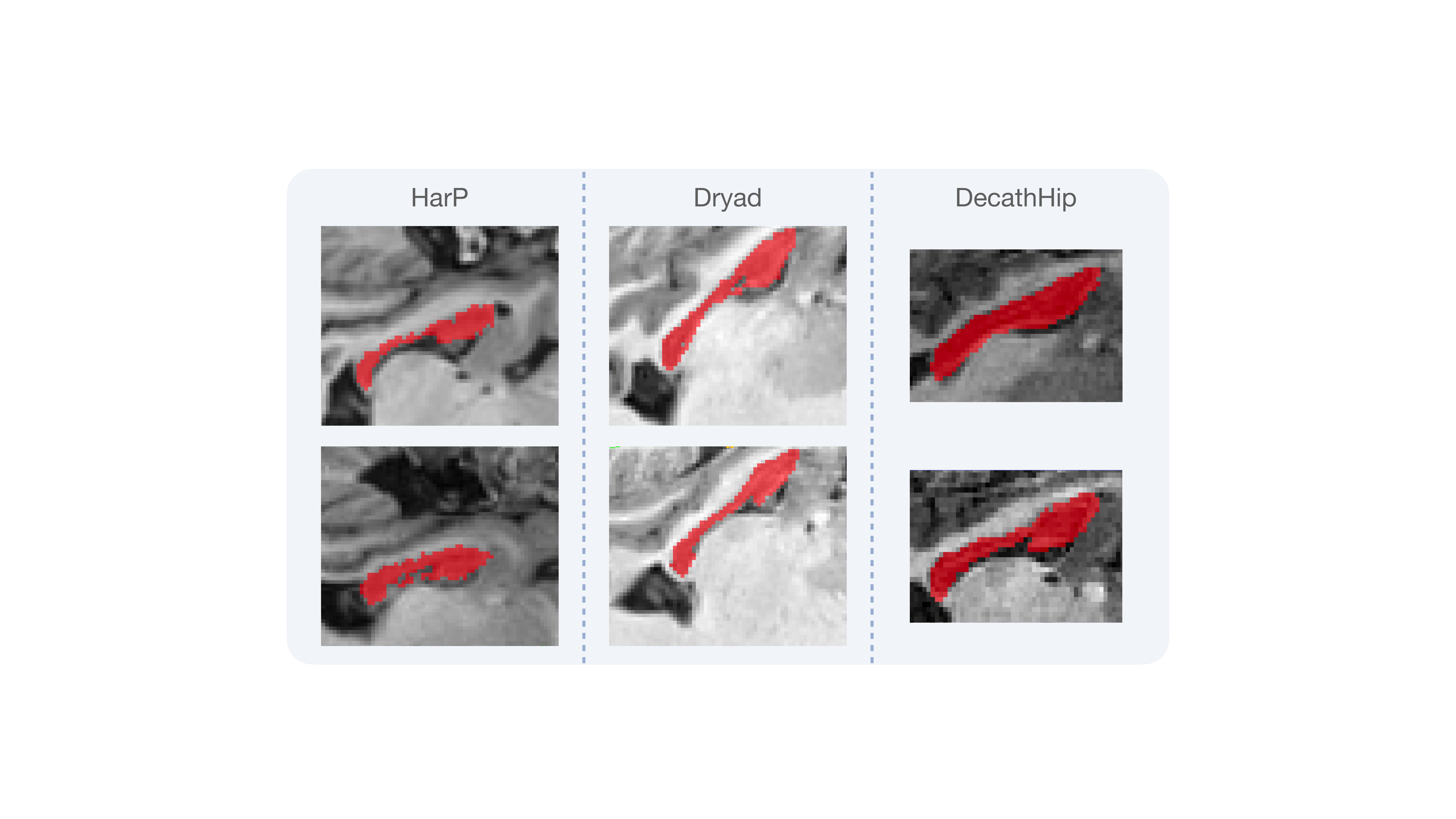}
\end{center}
   \caption{Samples from the hippocampus data with corresponding GT segmentations -- \cite{boccardi2015training, kulaga2015multi, medicaldecathlon}.}
\label{fig:hip_exs}
\end{figure}

Closely looking at the table shows that all three datasets have the same voxel spacing for all dimensions and nearly the same resolution. DecathHip has on average a smaller resolution than HarP or Dryad. Figure \ref{fig:hip_exs} shows samples from our data corpus with corresponding Ground Truth (GT) segmentations.

\subsection{Hardware settings} All experiments are conducted on a server system with 256GB DDR4 SDRAM, 2 Intel Xeon Silver 4210 CPUs and 8 NVIDIA Tesla T4 (16 GB) GPUs. To avoid any run time interference from multiple experiments on the same GPU, only one experiment is run on a single GPU at a time.

\subsection{Training Setup}
If not otherwise specified, every model is trained in a two-dimensional slice-by-slice model using the previously introduced hippocampus data corpus after being pre-processed by the nnU-Net framework. For further details on the pre-processing, the reader is referred to the original publication of the nnU-Net \cite{isensee2018nnu}.

A random 80:20 data split is used for training and testing. Models are trained sequentially always following the same order: Harp $\rightarrow$ Drayd $\rightarrow$ DecathHip. All networks are trained for 250 epochs on every dataset from the corpus, \ie 750 epochs in total. Further, the models are only trained once.

Whenever a setting is evaluated, the final model after training on the entire corpus is used to extract predictions. We use the S\o{}rensen–Dice (Dice) coefficient with corresponding standard deviation ($\sigma$) between patients as a measure of model performance.

\subsection{Hyperparameter Tuning}
To select the best hyperparameters, a total of three different settings are tested for every CL method after training 125 epochs per task. To this end, a random 80:20 split from the original training split is used. In case a method expects multiple hyperparameters, some parameters are fixed to a specific -- \textit{most commonly the default} -- value. Only those parameters are fixed where the outcome is somewhat predictable and the authors introduced a specific default value in their publication, \eg the $\alpha$ used to calculate the Fisher values in RWalk. Hyperparameters such as $\lambda$ in EWC have a huge impact on the overall performance and thus are never fixed but rather tuned.

\begin{table}[htb!]
\centering
\begin{adjustbox}{max width=\linewidth}
{\begin{tabular}{lccc}
\toprule 
\multirow{2}{*}{Method} & \multicolumn{3}{c}{Hippocampus} \\ \cmidrule{2-4}
 & Fixed params & Tuned params & Best params \\ \midrule \midrule
$\text{EWC}_{\text{nnU-Net},\ \mathcal{E}:\ *}$ & \multirow{4}{*}{--} & \multirow{4}{*}{$\lambda = \left\{ 0.01, 0.20, 0.50\right\}$} & $\lambda = 0.01$ \\
$\text{EWC}_{\text{ViT},\ \mathcal{E}:\ *}$ & &  & $\lambda = 0.01$ \\
$\text{EWC}_{\text{ViT},\ \mathcal{E}:\ \text{\scriptsize nnU-Net}}$ & &  & $\lambda = 0.01$ \\
$\text{EWC}_{\text{ViT},\ \mathcal{E}:\ \text{\scriptsize ViT}}$ & &  & $\lambda = 0.20$ \\ \midrule
$\text{RWalk}_{\text{nnU-Net}}$ & \multirow{2}{*}{\shortstack{$\alpha = 0.9,$ \\ update $= 10$}} & \multirow{2}{*}{$\lambda = \left\{ 0.80, 2.10, 3.30\right\}$} & $\lambda = 2.10$ \\
$\text{RWalk}_{\text{ViT}}$ & &  & $\lambda = 3.30$ \\ \midrule
$\text{MiB}_{\text{nnU-Net}}$ & \multirow{2}{*}{$\alpha = 0.9$} & \multirow{2}{*}{$\lambda = \left\{ 0.10, 1.00, 2.50\right\}$} & $\lambda = 1.00$ \\
$\text{MiB}_{\text{ViT}}$ & &  & $\lambda = 0.10$ \\ \midrule
$\text{POD}_{\text{nnU-Net}}$ & \multirow{2}{*}{scales $= 3$} & \multirow{2}{*}{$\lambda = \left\{ 0.01,0.10,0.20\right\}$} & $\lambda = 0.01$ \\
$\text{POD}_{\text{ViT}}$ & &  & $\lambda = 0.10$ \\ \midrule
$\text{PLOP}_{\text{nnU-Net}}$ & \multirow{2}{*}{scales $= 3$} & \multirow{2}{*}{$\lambda = \left\{ 0.01,0.10,0.20\right\}$} & $\lambda = 0.10$ \\
$\text{PLOP}_{\text{ViT}}$ & &  & $\lambda = 0.10$ \\
\bottomrule\\
\end{tabular}}
\end{adjustbox}
\caption{Overview of parameter setting considered for each CL method.}
\label{tab:ps}
\end{table}

Table \ref{tab:ps} summarises the parameter settings considered for  every method, where we regard the setting as best which achieves the highest mean Dice over all intermediate networks evaluated on all tasks. In case two parameter settings result in the same Dice, the one with the lowest $\sigma$ is used. This measure indirectly takes into account both the amount of forgetting and the Forward Transfer.

\section{Experimental Results}
We first conduct an ablation study to select the best performing ViT U-Net variation. In a second step, we analyse the relevance of the ViT when trained in a simple sequential setup, while freezing different components of the network. In a third set of experiments we further look into the behaviour and impact of different architecture components when applying the EWC regularisation term. Last but not least, we present the performance results of all considered CL methods in terms of backward and forward transfer.

Our experiments provide us with multiple valuable insights. Firstly, we give a clear understanding of the interplay between the U-Net and ViT within the architecture as well as their influence on each other. We also thoroughly demonstrate the importance and role of the self-attention mechanism in a CL setup. Furthermore, the relevance of the different architecture components is analysed in order to comprehend which are key for maintaining performance on previous tasks and which can be safely tuned for acquiring new knowledge. Within the context of a CL setup, this indicates what network modules can be safely regularised.

\subsection{Metrics}
Backward Transfer (BWT) indicates how much knowledge is maintained during training on a sequence of tasks and Forward Transfer (FWT) is used to demonstrate how a model state trained on $\{1, \dots, i-1\}$ performs on unseen, yet to train on dataset $i$. 

We define BWT for a specific task $\mathfrak{T}_{i}$ as
\begin{align}
\label{eqn:B}
    \text{BWT}\left( \mathfrak{T}_{i}\right) &=
    \text{Dice}\left(\mathcal{M}_{\left[ \mathfrak{T}_{1}, \dots, \mathfrak{T}_{i}, \dots, \mathfrak{T}_{n}\right]}, \mathfrak{T}_{i}\right) \nonumber\\
    &\:- \text{Dice}\left(\mathcal{M}_{\left[ \mathfrak{T}_{1}, \dots, \mathfrak{T}_{i}\right]}, \mathfrak{T}_{i}\right),
\end{align}

where $\mathcal{M}_{\left[ \mathfrak{T}_{1}, \dots, \mathfrak{T}_{k}\right]}$ is any network trained on data $\{1, \dots, k\}$ and $\text{Dice}(\mathcal{M}_{\left[ \mathfrak{T}_{1}, \dots, \mathfrak{T}_{j}\right]}, \mathfrak{T}_{i})$ indicates the performance from a network trained on data $\{1, \dots, j\}$ evaluated on dataset $i$. 

FWT for $\mathfrak{T}_{i}$ is correspondingly defined as

\begin{align}
\label{eqn:F}
    \text{FWT}\left( \mathfrak{T}_{i}\right) &= \text{Dice}\left(\mathcal{M}_{\left[ \mathfrak{T}_{1}, \dots, \mathfrak{T}_{i-1}\right]}, \mathfrak{T}_{i}\right) \nonumber \\
    &\:- \text{Dice}\left(\mathcal{M}_{\left[\mathfrak{T}_{i}\right]}, \mathfrak{T}_{i}\right),
\end{align}
based on \cite{diaz2018don, lopez2017gradient}. As Lopez-Paz and Ranzato mention in their work \cite{lopez2017gradient}, BWT for the first model state as well as FWT for the last model state are not defined.

Throughout this section, the Dice is used in multiple variations, where \emph{Dice mean} indicates the mean Dice of the final network evaluated on all three tasks, \emph{Dice first} indicates the Dice of the final network evaluated on the first task and \emph{Dice last} on the third task.

\subsection{Ablation study}
Shifted Patch Tokenization (SPT) and Locality Self-Attention (LSA) are ViT modifications to enable an efficient training of the ViT on small size datasets, introduced by Lee \etal~\cite{lee2021vision}. Combining our proposed ViT U-Net versions with the ViT modifications as suggested by Lee \etal~\cite{lee2021vision}, we end up with multiple ViT U-Net variations. To select the best-performing architecture, we first conduct an ablation study by analysing all variations for V1 and V2, \ie without using the proposed SPT or LSA method, using either one of those ViT adaptations or using both proposed modifications. We provide an overview of the results using the final model in Table \ref{tab:ablation}.

\begin{table}[htp]
\begin{center}
\begin{adjustbox}{max width=\linewidth}
{\begin{tabular}{lcccc}
\toprule
\multicolumn{2}{c}{\multirow{2}{*}{Ablation}} & \multicolumn{3}{c}{Dice $\uparrow{ } \pm{ } $ $\sigma \downarrow $ {[}\%{]}} \\ \cmidrule{3-5}
\multicolumn{2}{l}{} & HarP & Dryad & DecathHip \\ \midrule \midrule
\multirow{4}{*}{ViT U-Net V1} & \textit{unmodified} & $4.08 \pm{} 1.27$ & $3.82 \pm{} 0.76$ & $89.63 \pm{} 0.54$ \\ \cmidrule{2-5}
 & SPT & $4.79 \pm{} 1.35$ & $2.57 \pm{} 0.70$ & $89.71 \pm{} 0.55$ \\
 & LSA & $3.63 \pm{} 1.24$ & $5.49 \pm{} 0.52$ & $\mathbf{89.76 \pm{} 0.56}$ \\
 & SPT + LSA & $3.86 \pm{} 1.42$ & $6.35 \pm{} 0.38$ & $89.68 \pm{} 0.57$ \\ \midrule
\multirow{4}{*}{\textbf{ViT U-Net V2}}
 & \textit{unmodified} & $2.77 \pm{} 0.89$ & $3.70 \pm{} 0.51$ & $89.66 \pm{} 0.57$ \\ \cmidrule{2-5}
 & \textbf{SPT} & $\mathbf{4.82 \pm{} 1.83}$ & $\mathbf{16.19 \pm{} 0.94}$ & $89.69 \pm{} 0.55$ \\
 & LSA & $2.98 \pm{} 1.05$ & $6.19 \pm{} 0.43$ & $89.75 \pm{} 0.53$ \\
 & SPT + LSA & $3.88 \pm{} 1.09$ & $4.94 \pm{} 0.41$ & $89.67 \pm{} 0.61$ \\
\bottomrule
\end{tabular}}
\end{adjustbox}
\caption{Performance of the ViT U-Net ablations trained on the hippocampus data corpus without applying any CL mechanism.}
\label{tab:ablation}
\end{center}
\end{table}

The \textit{unmodified} rows indicate the unmodified ViT as opposed to the ViT variations (V1 or V2) and their combination (using any combination between SPT and LSA). We highlight the best method bold as well as the highest performance per task. The results show that the ViT U-Net V2 using only SPT and the base ViT architecture with $12$ layers and self-attention heads performs best when trained in a sequential setting on the hippocampus data corpus. It achieves the highest performances in most cases and already maintains a significant larger amount of knowledge compared to all other ablations. Therefore, this particular setting is used to conduct all further experiments.

\subsection{Impact of ViT in Sequential Training settings}

\begin{figure}[htp]
\begin{center}
\begin{subfigure}{0.49\linewidth}
  \includegraphics[trim={0 0 0 0.67cm},clip,width=\linewidth]{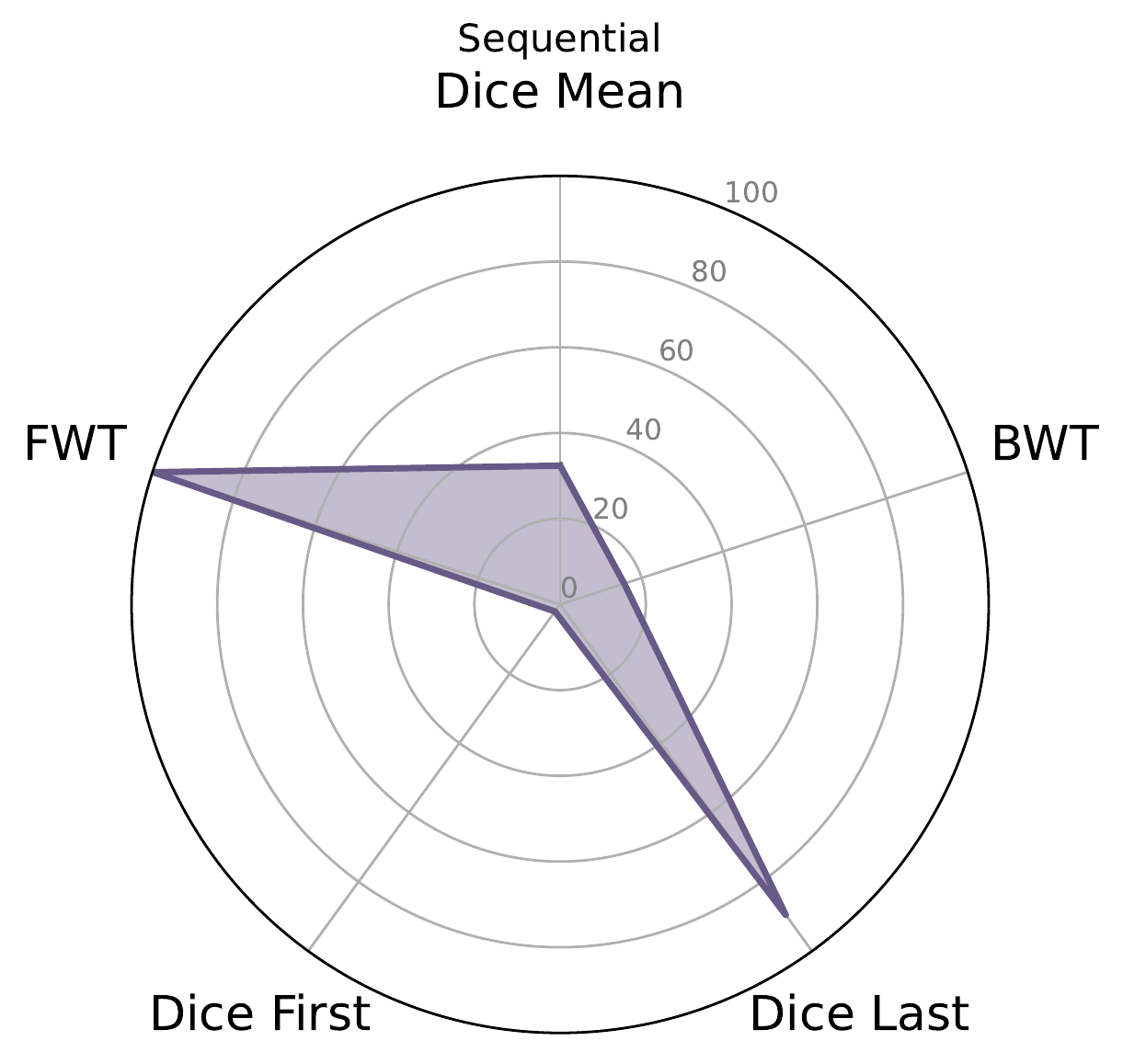}  
  \caption{Sequential nnU-Net}
  \label{fig:seq_unet}
\end{subfigure}
\begin{subfigure}{0.49\linewidth}
  \includegraphics[trim={0 0 0 0.67cm},clip,width=\linewidth]{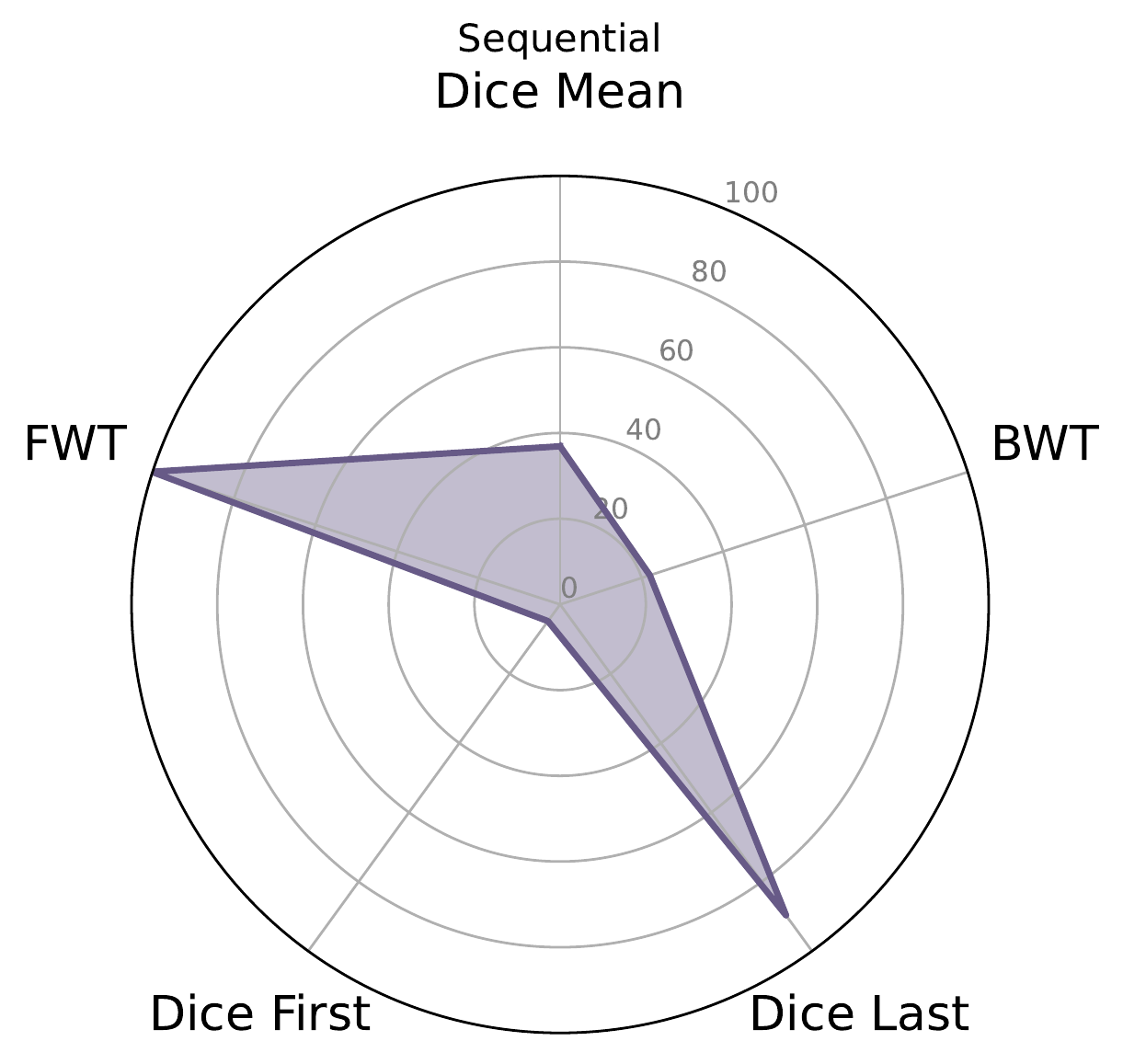}
  \caption{Sequential ViT U-Net}
  \label{fig:seq_vit}
\end{subfigure}

\begin{subfigure}{0.49\linewidth}
  \includegraphics[trim={0 0 0 0.6cm},clip, width=\linewidth]{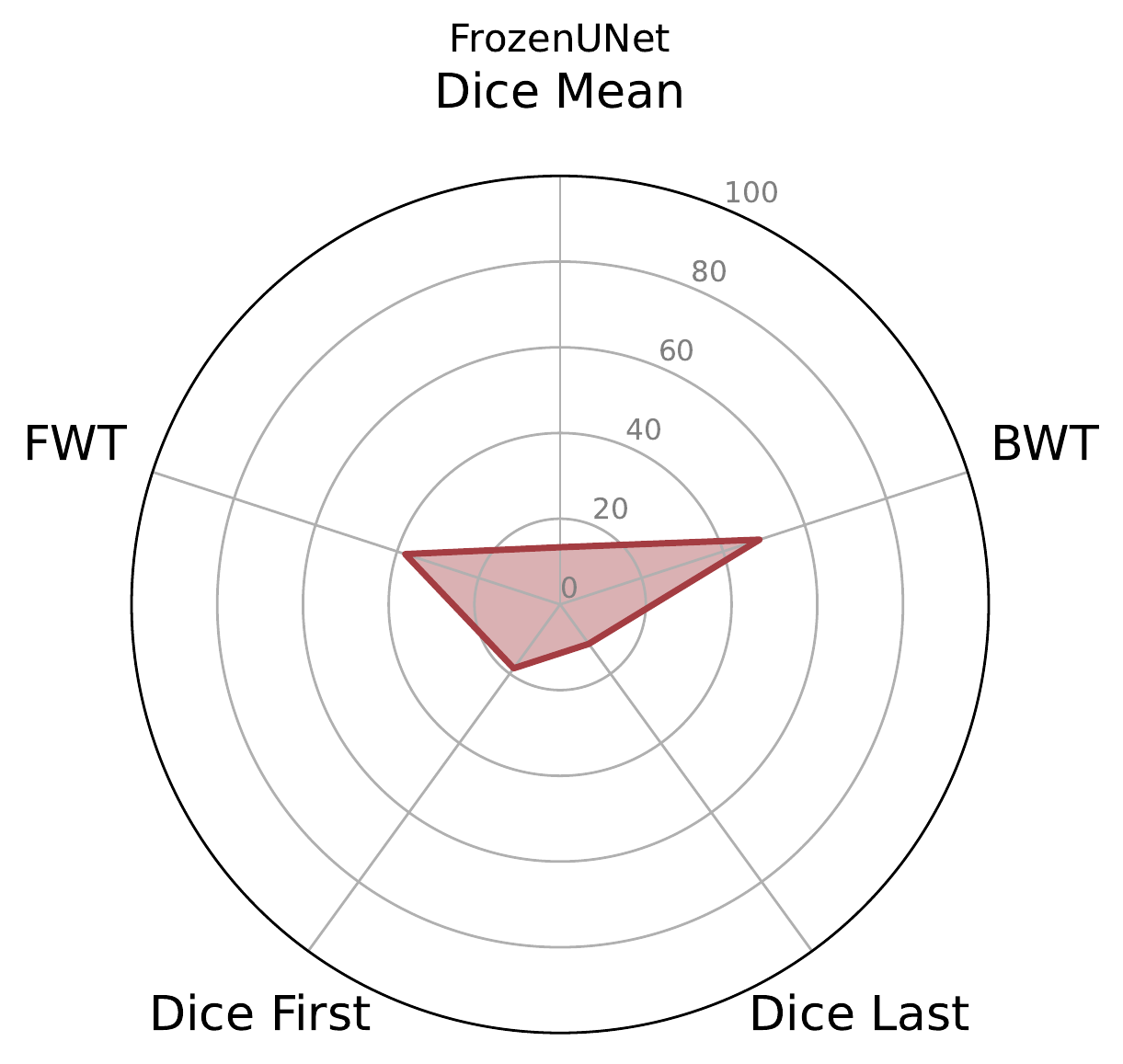}  
  \caption{Frozen U-Net}
  \label{fig:froz_un}
\end{subfigure}
\begin{subfigure}{0.49\linewidth}
  \includegraphics[trim={0 0 0 0.6cm},clip,width=\linewidth]{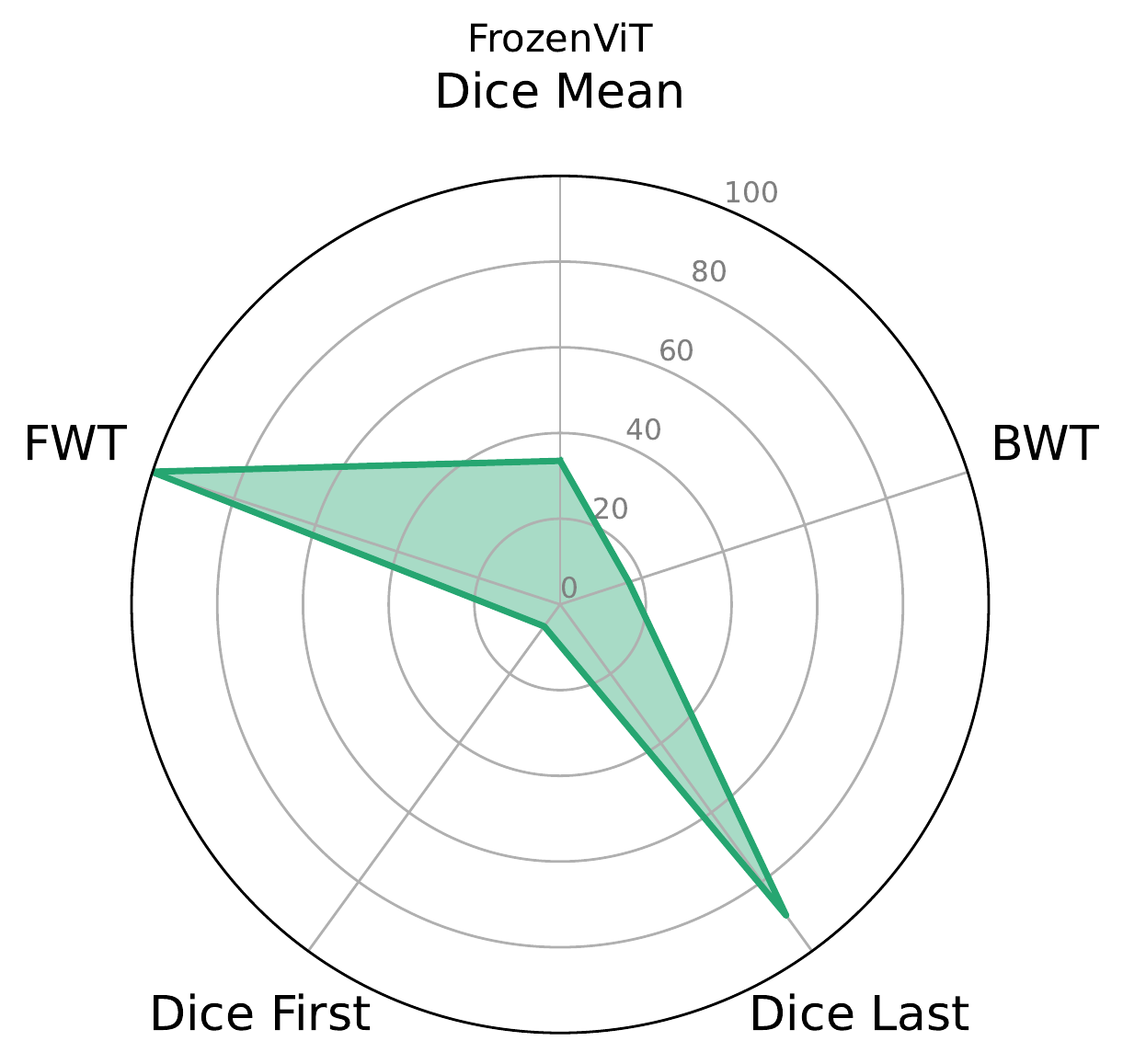}  
  \caption{Frozen ViT}
  \label{fig:froz_vit}
\end{subfigure}
\caption{Comparison of the network performances when freezing different parts of the network. No CL mechanism was used during training. The enclosed area is defined by mean Dice over all heads, the first and last Dice value as well as mean BWT and FWT over all tasks in $[\%]$. The larger the area, the better the performance of the method.}
\label{fig:frozen}
\end{center}
\end{figure}

To directly visualise if the self-attention mechanism helps to maintain more knowledge, the ViT U-Net architecture is used in two additional experiments. 
In the first, the U-Net is frozen after training on the first task -- a setting that we name \textit{Frozen U-Net} --. In a second experiment,  we freeze the ViT architecture instead after the first task, which we refer to as \textit{Frozen ViT}. Figure \ref{fig:frozen} shows the performance results of our four variations. For plotting purposes, we normalize the BWT and FWT values from the $\left[ -1; 1 \right]$ range into a positive $\left[0 ; 2 \right]$ interval by adding 1.0 to all values.

Comparing the two sequential training setups, a performance increase can easily be seen between the traditional nnU-Net (\ref{fig:seq_unet}) and ViT U-Net (\ref{fig:seq_vit}) architecture, as the amount of BWT is higher. Closely analysing the Frozen U-Net experiment (\ref{fig:froz_un}) one can deduct that knowledge from previous tasks is maintained the best as the BWT achieves the highest values. However, as the nnU-Net is frozen, the performance on the last task is very low and indicates that the nnU-Net components are crucial to learn new knowledge. The ViT architecture does not perform that good for medical image segmentation on its own. In comparison, the Frozen ViT variation (\ref{fig:froz_vit}) achieves very high results on the last task, however the performance on the first task is significantly lower indicating that the longer the ViT architecture is frozen, the more knowledge is forgotten. These findings confirm our expectations that  \textbf{the self-attention mechanism of the ViT can indeed be leveraged to maintain knowledge} when trained on a sequence of tasks for medical image segmentation.

\subsection{Impact of ViT during regularisation}
\begin{figure}[htp]
\begin{center}
\begin{subfigure}{0.49\linewidth}
  \includegraphics[trim={0 0 0 0.67cm},clip,width=\linewidth]{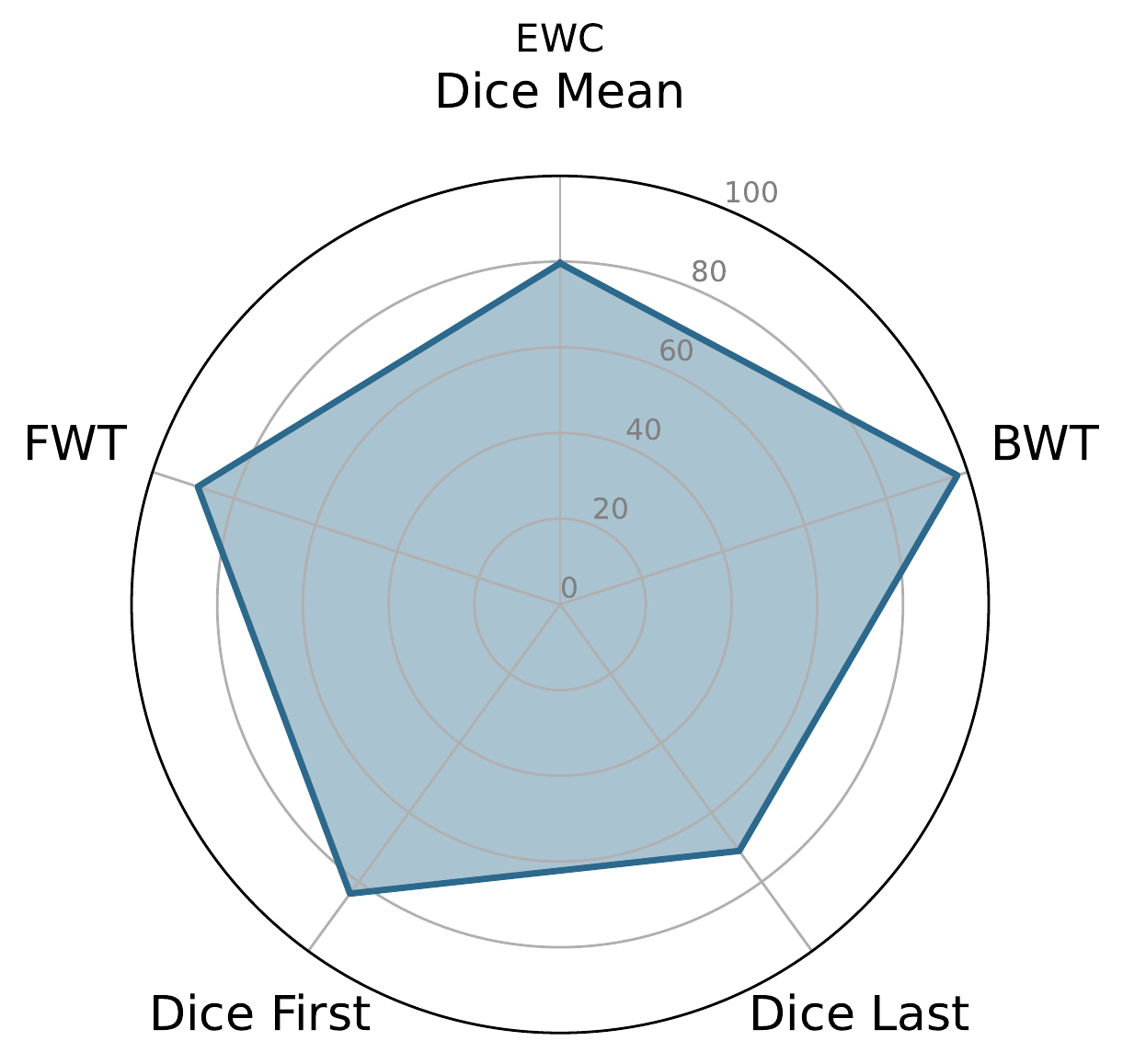}  
  \caption{EWC nnU-Net}
  \label{fig:ewc_unet}
\end{subfigure}
\begin{subfigure}{0.49\linewidth}
  \includegraphics[trim={0 0 0 0.67cm},clip,width=\linewidth]{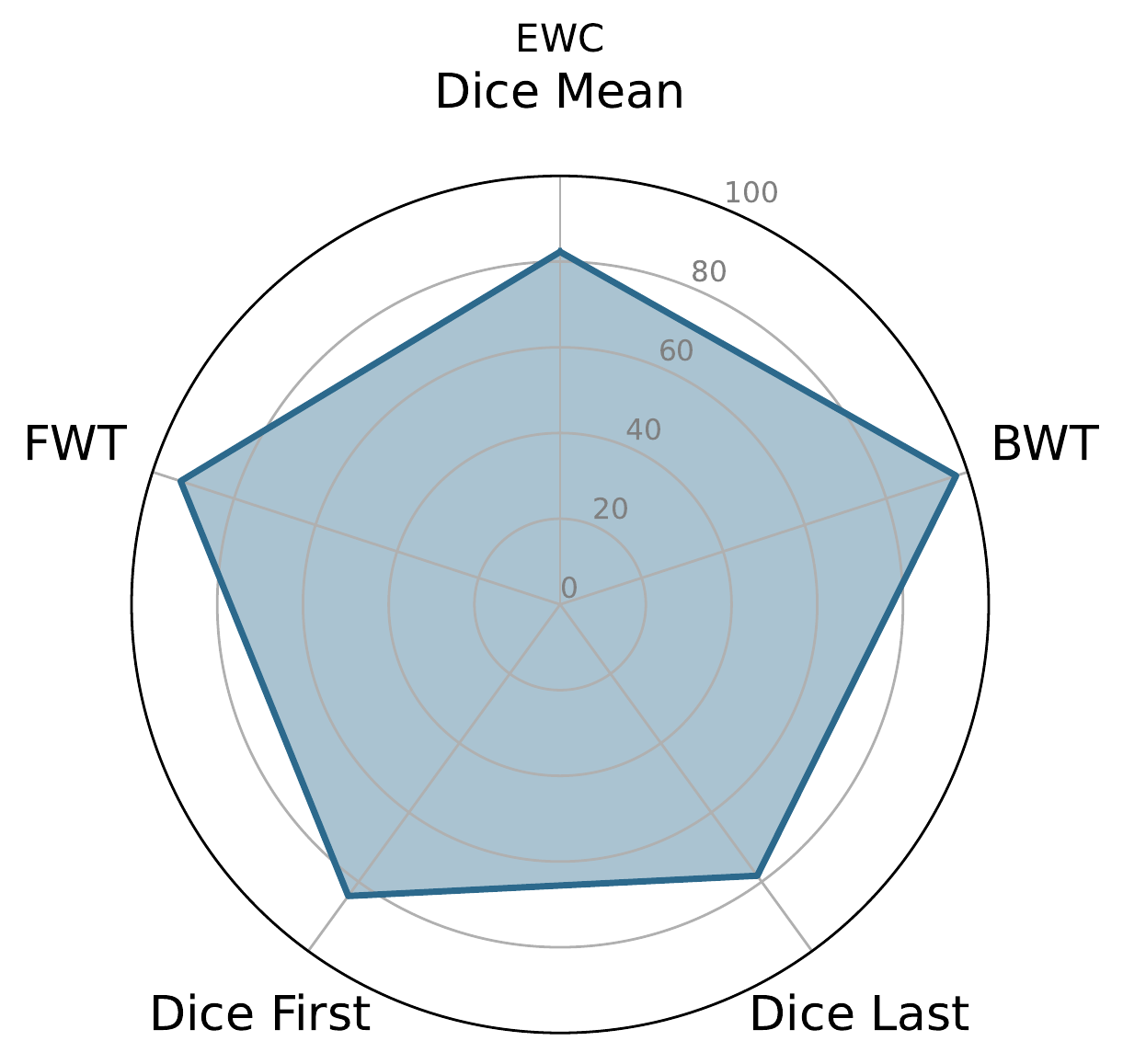}
  \caption{EWC ViT U-Net}
  \label{fig:ewc_vit}
\end{subfigure}

\begin{subfigure}{0.49\linewidth}
  \includegraphics[trim={0 0 0 0.6cm},clip, width=\linewidth]{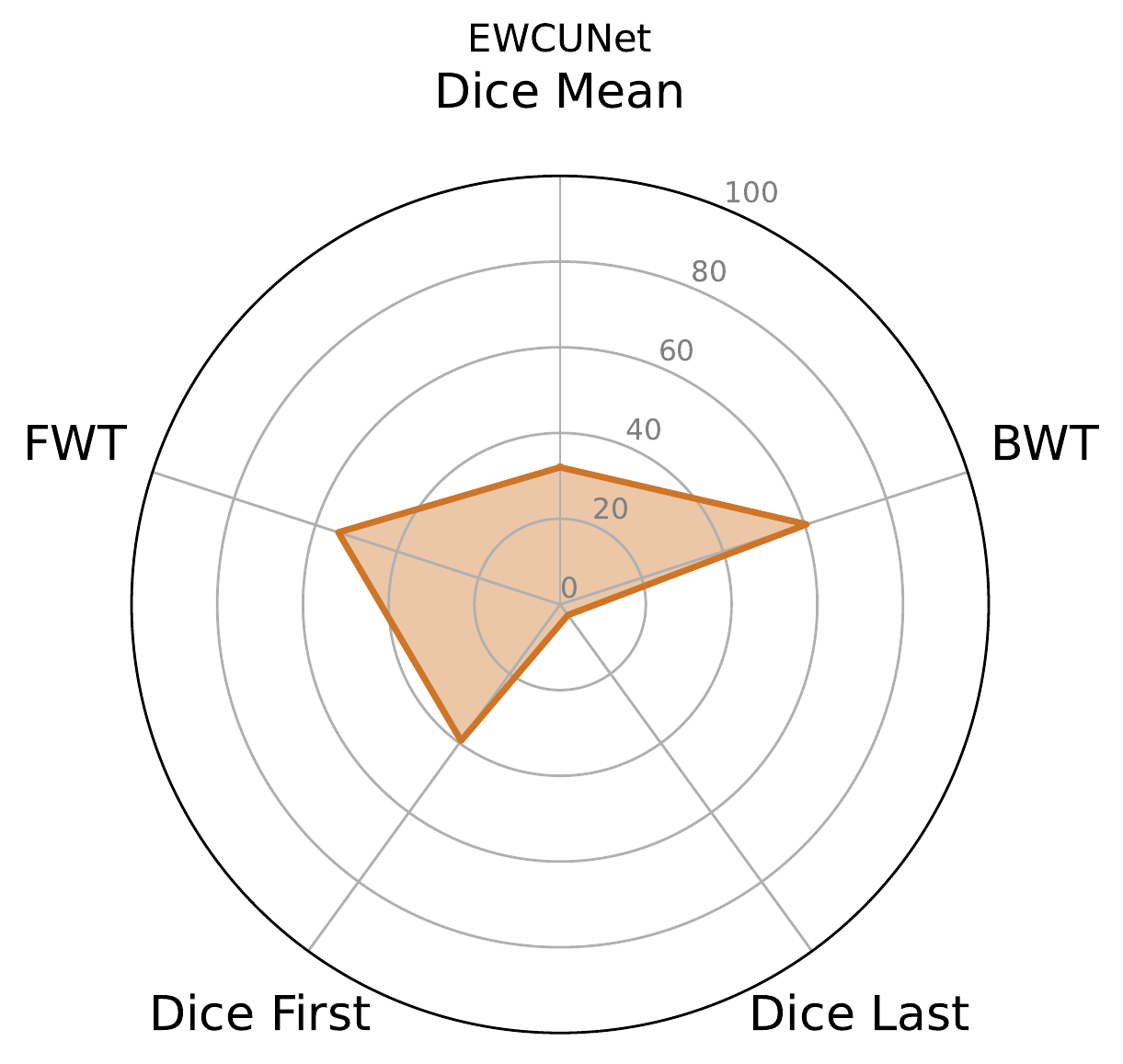}  
  \caption{EWC only on U-Net}
  \label{fig:ewc_o_unet}
\end{subfigure}
\begin{subfigure}{0.49\linewidth}
  \includegraphics[trim={0 0 0 0.6cm},clip,width=\linewidth]{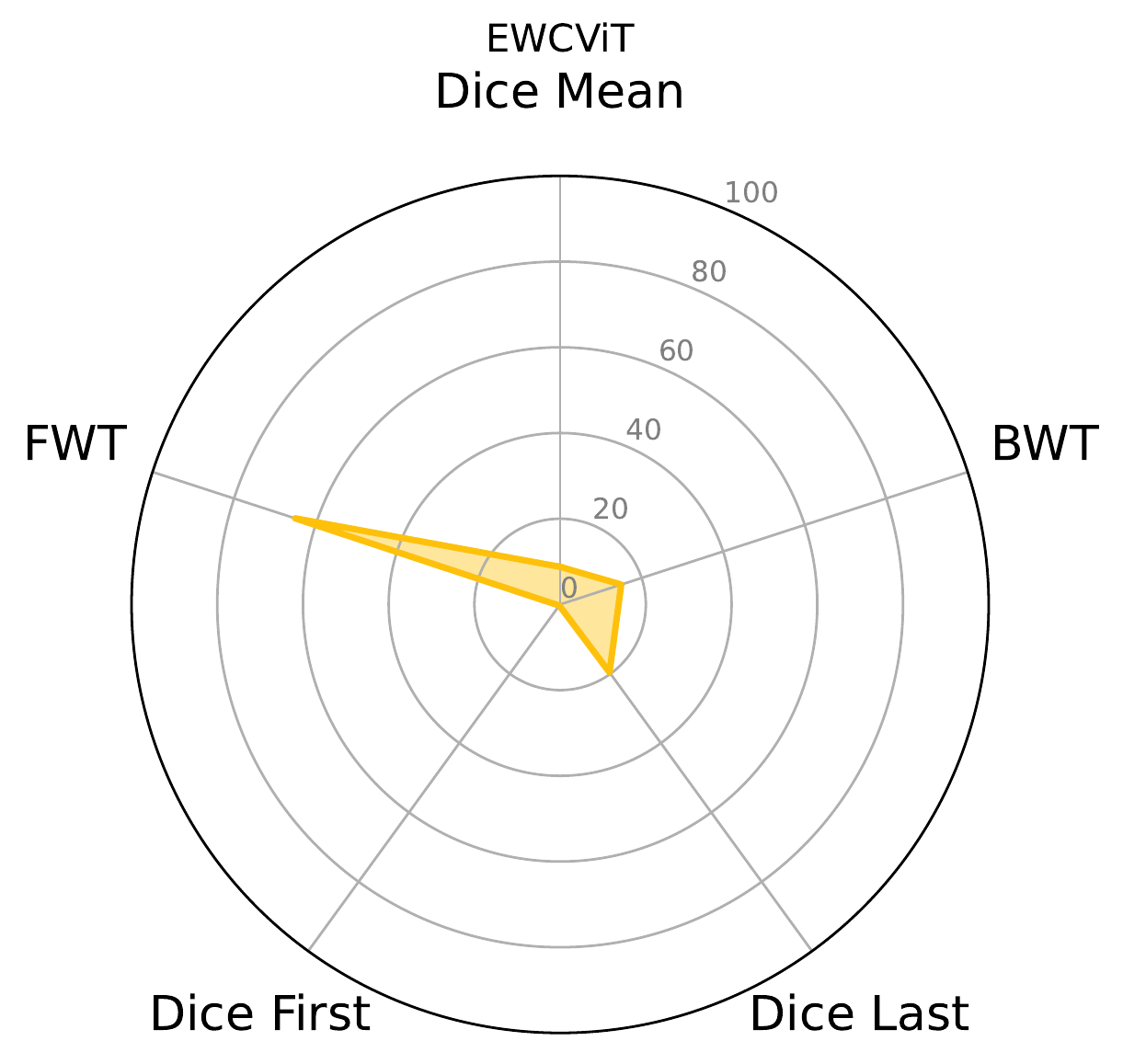}  
  \caption{EWC only on ViT}
  \label{fig:ewc_o_vit}
\end{subfigure}
\caption{Comparison of the network performances when applying EWC on different parts of the network. Enclosed area is defined by mean Dice over all heads, the first and last Dice value as well as mean BWT and FWT over all tasks in $[\%]$. The larger the area, the better the performance of the method.}
\label{fig:ewcs}
\end{center}
\end{figure}

In order to shed even more light on the role of the ViT in the network, we conduct four EWC-related experiments to demonstrate the influence of the architecture components and the effect of regularisation applied on different layers. Closely analysing Figure \ref{fig:ewcs}, one can see that the EWC on a plain nnU-Net (\ref{fig:ewc_unet}) already achieves very good results, as the enclosed area is large. However, using the ViT U-Net instead (\ref{fig:ewc_vit}), improves the overall performance. For instance, when FWT is increased, the Dice on the last task is enhanced as well, indicating that the ViT U-Net architecture maintains the amount of knowledge from previous tasks while achieving a higher Dice on the last trained task. Looking at the parameter settings from Table \ref{tab:ps}, the ViT U-Net uses the same $\lambda$ setting as the plain nnU-Net which clearly indicates that the EWC regularisation term has a softer impact on the ViT U-Net architecture, \ie decreasing the rigidity, as the learning on new tasks is fairly improved. 

Performing regularisation only on the ViT architecture (\ref{fig:ewc_o_vit}) leads to a much higher FWT value as well as higher performances on the last task as opposed to applying EWC only on the nnU-Net architecture (\ref{fig:ewc_o_unet}). However, only regularising the nnU-Net components leads to a significantly larger BWT as well as a better performance on the first task. Those findings indicate that \textbf{the regularisation on the ViT architecture interferes with the self-attention mechanism} and thus \textbf{decreases the amount of maintained knowledge}. Furthermore, this analysis shows that the composition of the nnU-Net with the ViT architecture creates some sort of interplay, \ie \textbf{convolutions and self-attention are complementary} as mentioned by Park and Kim in their recent work \cite{park2022vision}.

\subsection{Performance of CL methods}
All CL methods are deployed on the state-of-the-art nnU-Net to ensure a fair comparison. Table \ref{tab:res} shows an overview over all conducted experiments while reporting BWT and FWT and highlighting the differences in performance.

Analysing Table \ref{tab:res}, it is straightforward that the sequential, POD and PLOP training methods perform the worst. In fact, the POD and PLOP methods even perform worse than the sequential setting under certain circumstances. The rehearsal method embodies our upper bound and EWC achieves a performance which is more or less in the same range in terms of BWT and FWT. The Dice on the last task, however, is -- \textit{due to the rigidity of the regularisation} -- approximately $10 \%$ worse compared to the results of applying rehearsal. 

Surprisingly, RWalk performs much worse than EWC although it uses the same EWC regularisation as part of the overall regularisation method. This indicates that either the parameter importance scores worsens the regularisation or the online calculation of $\mathcal{F}$ does not represent a good alternative for medical imaging as opposed to the original calculation of $\mathcal{F}$ in EWC. On the other hand, the MiB method achieves -- \textit{excluding EWC and rehearsal} -- the best performances when only used with the nnU-Net. 

\begin{table}[htb!]
\centering
\begin{adjustbox}{max width=0.97\linewidth}
{\begin{tabular}{lccccc}
\toprule
\multirow{2}{*}{CL Method} & \multicolumn{2}{c}{$\text{BWT} \uparrow{ }$ {[}\%{]}} && \multicolumn{2}{c}{$\text{FWT} \uparrow{ }$ {[}\%{]}} \\ \cmidrule{2-3} \cmidrule{5-6}
  & HarP & Dryad && Dryad & DecathHip \\ \midrule \midrule

$\text{Sequential}_{\text{nnU-Net}}$ & $-83.47$ & $-85.09$ && $-0.22$ & $-0.08$ \\ 
$\text{Sequential}_{\text{ViT}}$ & $\mathbf{-81.03}$ & $\mathbf{-74.94}$ && $\;\;\;\mathbf{0.31}$ & $\;\;\;\mathbf{0.02}$ \\ \midrule
$\text{POD}_{\text{nnU-Net}}$ & $\mathbf{-82.80}$ & $-86.84$ && $\;\;\;0.53$ & $\;\;\;0.19$ \\
$\text{POD}_{\text{ViT}}$ & $-85.02$ & $\mathbf{-85.16}$ && $\mathbf{\;\;\;0.93}$ & $\;\;\;0.19$ \\ \midrule
$\text{PLOP}_{\text{nnU-Net}}$ & $\mathbf{-81.47}$ & $\mathbf{-82.64}$ && $\;\;\;\mathbf{0.37}$ & $-1.02$ \\ 
$\text{PLOP}_{\text{ViT}}$ & $-83.29$ & $-84.87$ && $\;\;\;0.23$ & $\mathbf{-0.92}$ \\ \midrule
$\text{RWalk}_{\text{nnU-Net}}$ & $\mathbf{-81.34}$ & $\mathbf{-83.12}$ && $-0.34$ & $-0.05$ \\ 
$\text{RWalk}_{\text{ViT}}$ & $-82.22$ & $-87.78$ && $\;\;\;\mathbf{0.43}$ & $\;\;\;\mathbf{0.01}$ \\ \midrule
$\text{MiB}_{\text{nnU-Net}}$ & $\mathbf{-82.61}$ & $\mathbf{-65.86}$ && $-0.15$ & $-0.89$ \\ 
$\text{MiB}_{\text{ViT}}$ & $-83.74$ & $-83.32$ && $\;\;\;\mathbf{0.30}$ & $\mathbf{-0.17}$ \\ \midrule
$\text{EWC}_{\text{nnU-Net}}$ & $-2.22$ & $\mathbf{-2.87}$ && $-3.65$ & $-18.57$ \\
$\text{EWC}_{\text{ViT}}$ & $\mathbf{-1.76}$ & $-3.95$ && $\mathbf{-2.39}$ & $\mathbf{-11.41}$ \\ \midrule
$\text{Rehearsal}_{\text{nnU-Net}}$ & $\mathbf{-0.23}$ & $\mathbf{-0.54}$ && $\;\;\;0.14$ & $-0.02$ \\ 
$\text{Rehearsal}_{\text{ViT}}$ & $-0.31$ & $-0.59$ && $\;\;\;\mathbf{0.44}$ & $\;\;\;\mathbf{0.20}$ \\
\bottomrule
\end{tabular}}
\end{adjustbox}
\caption{nnU-Net and ViT U-Net performance results for CL methods in form of BWT and FWT as defined in Equations \ref{eqn:B} and \ref{eqn:F}.}
\label{tab:res}
\end{table}

Only in some cases does the ViT U-Net actually decrease the amount of Catastrophic Forgetting. This undermines our previous finding, that regularising the ViT has a negative influence on the self-attention mechanism which might be an indicator for why the BWT is decreased in most of the cases. \textbf{If regularisation is not introduced on the ViT architecture as shown in the sequential training -- Table \ref{tab:ablation} and Figure \ref{fig:frozen} --, the results are significantly increased}. In most of the cases however, the \textbf{ViT U-Net achieves improved values for FWT}.

\begin{figure}[htp]
\begin{center}
\begin{subfigure}[t]{0.23\linewidth}
  \makebox[0pt][r]{\makebox[13pt]{\raisebox{140pt}{\rotatebox[origin=c]{90}{\footnotesize Sequential}}}}%
  \makebox[0pt][r]{\makebox[13pt]{\raisebox{80.5pt}{\rotatebox[origin=c]{90}{\footnotesize EWC}}}}%
  \makebox[0pt][r]{\makebox[13pt]{\raisebox{25pt}{\rotatebox[origin=c]{90}{\footnotesize MiB}}}}%
  \includegraphics[trim={0cm 0 0cm 0cm},clip,width=\linewidth]{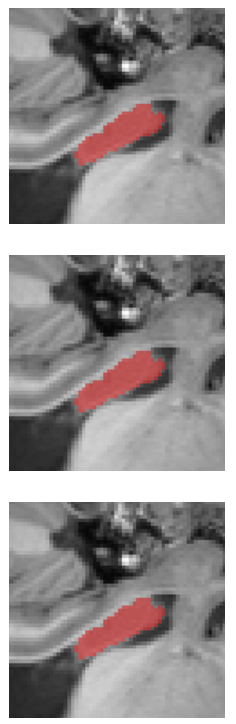}
  \caption{Image with GT}
  \label{fig:GT}
\end{subfigure}
\begin{subfigure}[t]{0.23\linewidth}
  \includegraphics[trim={0cm 0 0cm 0cm},clip,width=\linewidth]{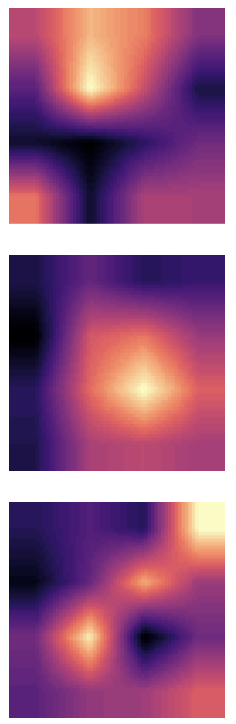}
  \caption{Attention heat map}
  \label{fig:AHM}
\end{subfigure}
\begin{subfigure}[t]{0.23\linewidth}
  \includegraphics[trim={0cm 0 0cm 0cm},clip,width=\linewidth]{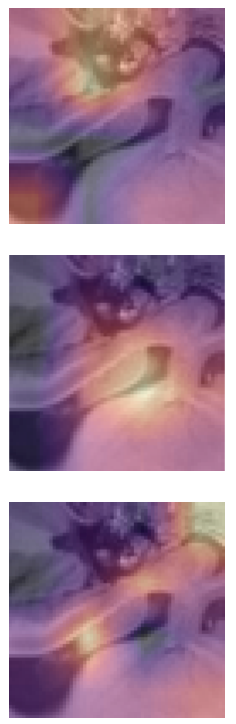}
  \caption{Image with attention heat map}
  \label{fig:overlay}
\end{subfigure}
\begin{subfigure}[t]{0.23\linewidth}
  \includegraphics[trim={0cm 0 0cm 0cm},clip,width=\linewidth]{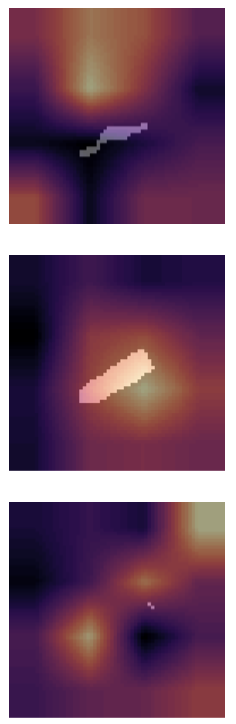}
  \caption{Prediction with attention heat map}
  \label{fig:pred}
\end{subfigure}

\caption{Analysis of the self-attention from the last ViT head using the intermediate networks trained on HarP and Dryad in a sequential (top row), EWC (middle row) and MiB (bottom row) setting using a random HarP image.}
\label{fig:trans_res}
\end{center}
\end{figure}

Finally, Figure \ref{fig:trans_res} shows an example prediction on a random HarP scan using ViT U-Nets trained only on HarP and Dryad in a sequential and EWC setting to showcase the ViT's attention and correspondingly influence on the network's prediction. For this purpose, the attention map from the last ViT head is used, \ie head 12 as we only use the base ViT architecture in the scope of this manuscript. The Figure also shows the predictions of the networks to allow a comparison with the GT segmentation.

The prediction of the sequential training method (\ref{fig:pred}) shows that a certain amount of forgetting has already taken place even though the network has not even started on task three yet. Comparing EWC with sequential training, one can certainly confirm that the EWC predictions match the GT (\ref{fig:GT}) quite well. Focusing on the attention heat maps (\ref{fig:AHM}) and corresponding overlay with the original image (\ref{fig:overlay}), one can easily see that the attention of the sequential network is off, \ie the attention is not even close to the area that should actually be segmented. This showcases that a significant amount of forgetting has taken place throughout the training on the second task.

Looking at the results from the MiB network, one can hardly find the network's prediction as it is a very small area. Interestingly, the main attention is located at the right top corner of the image, whereas parts of the attention are located over the GT area. This visually shows that the ViT has an influence on the prediction based on its focus within the image. If the focus is on a different area, the segmentation result is not good either, however this does not mean that the attention has a direct influence on the network's prediction as shown in the sequential setup. This shows that the nnU-Net component still has a larger influence on the prediction's shape and location within the image.

Shifting to the network trained in an EWC setting, the attention of the last ViT head focuses adequately on the hippocampus area where the GT is defined. However, the main focus is still a little off and not in the middle of the actual area that should be segmented. Nevertheless, the area of focus certainly captures the GT shown in Figure \ref{fig:GT}.

\section{Limitations}
Despite several insightful takeaways, our study has certain limitations. The first is our reduced scope, as we only focus on domain incremental learning and one anatomy for our analysis due to limited computational resources. It would be very interesting to know the differences and results when applied in a class incremental setup instead.

As the ViT U-Net is a large network with many parameters due to the self-adapting process, the amount of allocated resources are increased as compared to the traditional nnU-Net. Replacing convolutional layers with convolutional block attention modules for example might have a positive influence on the overall performance while reducing the amount of parameters \cite{yang2022aa}. Investigating different optimisations to lighten the amount of resource allocation and with it the overall ViT U-Net architecture is an interesting topic on its own.

Another rather intriguing aspect for future work is the importance and role of Layer Norms within the ViT, as recent work has shown that Batch-Norm Layers play an important role in terms of Catastrophic Forgetting \cite{gupta2020unreasonable, gupta2021addressing}. Within this scope, it would be interesting to see the outcome when replacing the ViT Layer Norms with Batch-Norm Layers, how the architecture responds if the ViT Norm Layers are frozen or even regularised using EWC or any suitable knowledge distillation.

\section{Conclusion}
As recent publications show, Transformers are taking a bigger role in the fields of medical imaging and semantic segmentation. With the ViT U-Net we successfully incorporate the ViT into the well-known and commonly used nnU-Net framework. Our ablation study shows that the primary task of image segmentation performance is not jeopardised by the combination of both architectures.

Additionally, we demonstrate that the proposed model leverages the self-attention of Transformers to maintain knowledge and thus mitigates Catastrophic Forgetting for medical imaging compared to the purely convolutional nnU-Net. 

However, regularising the ViT component over time has a negative effect in terms of maintaining knowledge, as regularisation interferes with the self-attention mechanism. Instead, discouraging the amount of change for convolutional layers is preferable for maintaining previous knowledge without compromising plasticity.

Given our promising findings, we hope that other researchers are inspired to conduct further research within the scope of Transformers in CL and medical image segmentation.

\section{Reproducibility}
All datasets used in this manuscript are publicly available under the corresponding citations. The random data split and trained networks along with instructions on how to run all experiments will be provided upon acceptance. The code including all implementations is publicly available under \url{https://github.com/MECLabTUDA/Lifelong-nnUNet}.

\twocolumn[{\centering{\Large \bf Supplementary Material \par}\vskip .5em\vspace*{21pt}}]

\setcounter{section}{0}

This supplementary material is split into two components. We first present the state-of-the-art when it comes to Transformer architectures and their application for medical image segmentation. This is used in addition to the main manuscript to put the main manuscript in context with existing work. Secondly, we include a simple, non-CL comparison between the no new Net (nnU-Net) and our proposed architecture from the manuscript.

\section{Related Work}
Transformers as presented in Vaswani \etal~\cite{vaswani2017attention} achieved huge success in the fields of Natural Language Processing (NLP) and Machine Translation over the last couple of years due to their ability of handling sequential input data by using self-attention \cite{brown2020language, devlin2018bert, khan2021transformers}.


\subsection{Transformer for medical image segmentation}
When it comes to medical image segmentation, the most common architecture used is the U-Net \cite{ronneberger2015u} or different variations like U-Net++ \cite{zhou2018unet++}, no new U-Net (nnU-Net) \cite{isensee2018nnu} or Deep Residual U-Net introduced by Zhang \etal~\cite{zhang2018road}. It is only recently that architectures for medical image segmentation relying solely on Transformer architectures or hybrid approaches have been presented.

Karimi \etal~\cite{karimi2021convolution} introduce a medical image segmentation network using Transformers instead of a Convolutional Neural Network (CNN). The presented architecture however is very similar to the introduced Vision Transformer (ViT) \cite{vaswani2017attention}, except it is developed for three-dimensional data like Computer Tomography (CT) scans. The authors show that the positional embedding makes a significant difference in terms of segmentation accuracy.

Another application of Transformer architectures -- \textit{TransBTS} -- for medical image segmentation is proposed by Wang \etal~\cite{wang2021transbts} with the specific use for multimodal brain tumor segmentation in Magnetic Resonance Images (MRIs). The presented approach is based on a three-dimensional CNN encoder -- decoder combined with a Transformer encoder in between. The Transformer encoder has the same architecture as the ViT.


\subsection{Hybrid U-Net Transformer architectures for medical image segmentation}
\textit{TransUNet}, presented by Chen \etal~\cite{chen2021transunet}, is a hybrid network that combines a CNN-Transformer hybrid model with the conventional U-Net architecture. The authors use CNN in order to extract features and to create a feature map. Regarding the input of the Transformer encoder, patches are extracted from the CNN feature maps instead of the raw input images to increase the performance of the architecture \cite{chen2021transunet}. The encoder is followed with a cascaded upsampler to predict the final segmentation by using multiple upsampling steps.

The proposed \textit{UNETR} \cite{hatamizadeh2021unetr} follows very closely the structure of the ViT architecture \cite{dosovitskiy2020image}. Three-dimensional volumes are split into three-dimensional patches that are linearly projected and flattened. Those patches are then fed into the Transformer network, whereas different encoded representations from Transformer layers are combined with the decoder using skip connections for predicting the segmentation mask. All in all, the U-Net encoder is replaced with the Transformer encoder and connected to the upsampling decoder that is then used to predict the final segmentation. The authors evaluation has shown that the presented method outperforms the TransUNet, TransBTS, but also baseline architectures like the nnU-Net.

\section{nnU-Net and ViT U-Net comparison in a non-CL setup}
In this supplementary part we provide the evaluation results of plain nnU-Nets and ViT U-Nets in a non-CL setup to compare their performances. For this purpose, we train and evaluate for every dataset from the hippocampus corpus one nnU-Net and one ViT U-Net respectively while evaluating the final network on all three hippocampus datasets. Table \ref{tab:res} shows the results based on the Dice scores. Bold values indicate the highest score between the nnU-Net and ViT U-Net. The same evaluation and experimental setups as explained in the main manuscript apply here as well.

\begin{table}[htb!]
\centering
\begin{adjustbox}{max width=0.97\linewidth}
{\begin{tabular}{lcccc}
\toprule
\multirow{2}{*}{Trained on} & \multirow{2}{*}{Architecture} & \multicolumn{3}{c}{$\text{Dice} \uparrow{ } \pm{} \sigma \downarrow{ }$ {[}\%{]}} \\ \cmidrule{3-5}
  & & HarP & Dryad & DecathHip \\ \midrule \midrule
\multirow{2}{*}{HarP} & nnU-Net & $85.72 \pm{} 0.77$ & $\mathbf{84.96 \pm{} 0.22}$ & $1.27 \pm{} 0.24$ \\
 & ViT U-Net & $\mathbf{85.74 \pm{} 0.99}$ & $84.81 \pm{} 0.24$ & $\mathbf{1.59 \pm{} 0.38}$ \\ \midrule
\multirow{2}{*}{Dryad} & nnU-Net & $\mathbf{38.76 \pm{} 5.26}$ & $90.82 \pm{} 0.27$ & $7.18 \pm{} 1.54$ \\
 & ViT U-Net & $34.63 \pm{} 7.10$ & $\mathbf{90.96 \pm{} 0.48}$ & $\mathbf{8.01 \pm{} 2.52}$ \\ \midrule
\multirow{2}{*}{DecathHip} & nnU-Net & $\mathbf{3.69 \pm{} 1.20}$ & $18.31 \pm{} 1.65$ & $89.67 \pm{} 0.40$ \\
 & ViT U-Net & $3.67 \pm{} 1.51$ & $\mathbf{20.13 \pm{} 0.98}$ & $\mathbf{89.69 \pm{} 0.39}$ \\
\bottomrule
\end{tabular}}
\end{adjustbox}
\caption{nnU-Net and ViT U-Net comparison in terms of performance results based on Dice scores.}
\label{tab:res_supp}
\end{table}

Briefly analysing Table \ref{tab:res_supp} it is easy to see that the ViT U-Net architecture outperforms the nnU-Net in two out of three times for every trained dataset. As the ViT U-Net architecture is not the main focus of the manuscript, we do not further analyse this comparison. However, it is worth mentioning that the performance differences are not significant enough to decide which architecture performs best in a non-CL setup.

{\small
\bibliographystyle{ieee_fullname}
\bibliography{Literature/literature.bib}
}

\end{document}